\documentclass[aps,floats,11pt, nofootinbib,prd,showkeys]{revtex4-2}

\usepackage{graphicx}
\usepackage{dcolumn}
\usepackage{bm}

\usepackage{epsfig}
\usepackage{epstopdf}
\usepackage[latin1]{inputenc}
\usepackage{amsmath}
\usepackage{hyperref}
\usepackage{xcolor}

\def\beq{\begin{equation}}
\def\eeq{\end{equation}}

\def\beqstr{\begin{equation*}}
\def\eeqstr{\end{equation*}}

\def\ber{\begin{eqnarray}}
\def\eer{\end{eqnarray}}

\def\berstr{\begin{eqnarray*}}
\def\eerstr{\end{eqnarray*}}

\def\benu{\begin{enumerate}}
\def\eenu{\end{enumerate}}

\def\l{\left}
\def\r{\right}

\def\f{\frac}

\begin{document}

\title{The effective speed of sound in cosmological perturbation theory}%

\author{Sanil Unnikrishnan}
\email{u.sanil@ststephens.edu}

\affiliation{Department of Physics, St.\ Stephen's College, University of Delhi, Delhi 110007, India}

\date{04 July 2026}
            
\begin{abstract}
In a multi-field/fluid cosmological system consisting of minimally coupled canonical scalar fields, non-canonical scalar fields, and barotropic perfect fluids, we introduce a new definition for the effective speed of sound of the entire system to describe the evolution of cosmological perturbations. This effective speed of sound is not only gauge-invariant but also a background-dependent quantity; it can, therefore, be treated as a parameter to quantify perturbations in such multi-field/fluid systems. It is with this effective speed that the gauge-invariant Bardeen potential and the curvature perturbation propagate at scales much smaller than the sound horizon. Furthermore, the effective speed of sound defined in this paper generalizes the definition provided by Garriga and Mukhanov for a single non-canonical scalar field to a system consisting of multiple minimally coupled barotropic perfect fluids, canonical scalar fields, and non-canonical scalar fields. Moreover, as in the case of a single pure-kinetic non-canonical scalar field, this effective speed of sound for the total system turns out to be identically equal to the total adiabatic speed of sound when the dynamics of the universe are driven by multiple pure-kinetic non-canonical scalar fields. This makes such a system tantamount to a system of equivalent multi-barotropic perfect fluids. We also derive a set of equations governing the evolution of perturbations in a general multi-field/fluid universe. Using these equations, we demonstrate that in the large-scale limit  ($k \to 0$), if the perturbations are initially adiabatic, they remain so at those scales throughout the evolution of the universe, thus extending this well-known result to a general multi-field/fluid system consisting of non-canonical scalar fields. Consequently, at those scales, such a multi-field/fluid universe dynamically behaves as if it contains only a single barotropic perfect fluid.
\end{abstract}

\maketitle
\flushbottom

\section{\label{Sec: introdiction}Introduction}

Within the framework of Einstein's general theory of relativity, the current understanding of the evolution of the universe, from the inflationary epoch to the present accelerated expansion phase, is reasonably well established and is in concordance with observational data~\cite{Planck:2015sxf, Planck:2018nkj, Planck:2018vyg, SupernovaSearchTeam:1998fmf, SupernovaCosmologyProject:1998vns, DESI:2024mwx, DiValentino:2020evt}. The expansion of the universe is driven by different components at various stages of its evolution. Scalar fields drive the inflationary phase~\cite{Starobinsky:1980te,Guth:1980zm,Linde:1981mu,Albrecht:1982wi,Linde:1983gd,Bassett:2005xm,Baumann:2009ds,Sriramkumar:2009kg,Mishra:2024axb}, followed by radiation and cold dark matter dominated expansion phases, which can be described as barotropic perfect fluids. 
By definition, barotropic perfect fluids are those whose pressure is a function of energy density alone, both at the background level and at all orders of perturbations~\cite{Unnikrishnan:2010ag, Arroja:2010wy}. The nature of dark energy~\cite{Sahni:2004ai,Copeland:2006wr,Li:2011sd}, which drives the late-time accelerated expansion of the universe, is not yet fully understood; however, many models, such as the cosmological constant and scalar field dark energy, remain viable according to current observational data~\cite{Sahni:1999gb, Peebles:2002gy, Carroll:2000fy, Padmanabhan:2002ji, Tsujikawa:2013fta, Frieman:2008sn}. 

Scalar fields with a standard Lagrangian are classified as canonical scalars, whereas those with a Lagrangian that is a general function of the kinetic term $X = \frac{1}{2}\partial_\mu\phi\partial^\mu\phi$ and the scalar field $\phi$ are classified as non-canonical scalars~\cite{Garriga:1999vw}. Non-canonical scalar fields have been widely studied to model inflation~\cite{Armendariz-Picon:1999hyi,Chen:2006nt} and as dark energy candidates driving the late-time accelerated expansion of the universe~\cite{Armendariz-Picon:2000ulo, Armendariz-Picon:2000nqq}. 

Unlike canonical scalar field models of inflation with monomial power-law potentials of the form $V(\phi) = V_{0}\phi^n$, non-canonical scalar field models with $\mathcal{L} = X^\alpha - V_{0}\phi^n$ (where $\alpha$ and $n$ are constants) can lead to observationally viable values for the scalar spectral index and the tensor-to-scalar ratio for integral values of $n$, such as $n = 2$ and $n = 4$~\cite{Unnikrishnan:2012zu, Li:2012vta}. This is possible in non-canonical scalar field models because of their smaller speed of sound, which consequently lowers the tensor-to-scalar ratio to a value well within observational bounds~\cite{Mishra:2022ijb, Unnikrishnan:2013vga, Rezazadeh:2014fwa}. Furthermore, non-canonical scalar fields can drive the late-time accelerated expansion of the universe~\cite{Chiba:1999ka} and provide a unified description of dark matter and dark energy~\cite{Scherrer:2004au, Sahni:2015hbf, Mishra:2018tki}. However, it is impossible to uniquely determine the fundamental nature of the non-canonical Lagrangian merely from cosmological observations, as two different non-canonical scalar field models can lead to the same observational effects~\cite{Unnikrishnan:2008ki}. 

For each type of matter content in the universe, one can define an equation of state parameter $w$ as the ratio of the pressure to the energy density. It is also possible to define the equation of state parameter for the total system consisting of various types of matter. Once the equation of state parameter and its evolution are known, the rate of expansion of the universe can be determined from the Friedmann equations. In addition to $w$, the effective speed of sound of the matter content is an important parameter that determines the evolution of perturbations in cosmological perturbation theory. For barotropic perfect fluids, the effective speed of sound is related to the equation of state parameter, whereas for canonical scalars, it is identically equal to unity~\cite{Hu:1998kj}. For non-canonical scalars, the effective speed of sound can take any value depending on the form of the Lagrangian~\cite{Garriga:1999vw}. In a multi-fluid/field system, although it is possible to define the effective speed of sound for each matter component separately, it is not yet established how one can define an effective speed of sound for the total system in a similar manner.

In cosmological perturbation theory, simply defining the square of the speed of sound as the ratio of the pressure perturbation $\delta p$ to the energy density perturbation $\delta \rho$ does not work in all cases. This is because both $\delta p$ and $\delta \rho$ are gauge-dependent quantities; in the case of scalar fields, the gauge in which $\delta \rho = 0$ does not, in general, coincide with the gauge in which $\delta p = 0$. Consequently, the square of the speed of sound defined as $\delta p / \delta \rho$ can become infinite in certain gauges, making such a definition ambiguous. Since the speed of sound must depend on the physical properties of the matter rather than the choice of gauge, one typically defines it as the ratio of a gauge-invariant pressure perturbation to the corresponding gauge-invariant energy density perturbation~\cite{Unnikrishnan:2008ki}. The speed of sound thus defined for a single scalar field or a barotropic perfect fluid is not only gauge-invariant but also a background-dependent quantity.

In the case of a multi-field system, it is possible to define an effective speed of sound as the ratio of the pressure perturbation to the energy density perturbation in the comoving gauge~\cite{Romano:2018frb,Rodrguez:2020hot,Romano:2023bzn}. In general, it turns out that this effective speed of sound is a space-dependent quantity~\cite{Romano:2018frb}. Such a definition of the effective speed of sound may be advantageous for expressing cosmological perturbations in systems consisting of non-minimally coupled matter components, as well as in modified gravity models. 

In this paper, we consider a cosmological system consisting of several minimally coupled matter components, such as barotropic perfect fluids, canonical scalar fields, and non-canonical scalar fields. In such a system, we introduce a new definition for the total effective speed of sound of a multi-field/fluid universe, based on the principle that the relations between quantities describing matter perturbations must be independent of the gauge choice. The effective speed of sound emerges as a parameter in a gauge-invariant relation between pressure and energy density perturbations; we demonstrate that, at small scales, the metric perturbations propagate at this specific speed. Furthermore, it is shown that in such a minimally coupled multi-field/fluid system, this effective speed of sound is not only gauge-invariant but also space-independent (or $k$-independent) and background-dependent (i.e., it depends on time only). Being a background-dependent quantity, it can be treated as a parameter to quantify cosmological perturbations in such a multi-field/fluid system.

This paper is organized as follows. In the following section, we briefly review first-order cosmological perturbation theory with scalar degrees of freedom. All the quantities required to describe scalar perturbations, their corresponding gauge-invariant definitions, and the equations governing their evolution are described in that section. In Sec.~\ref{Sec: single fluid/field}, we discuss the evolution of cosmological perturbations when the dynamics of the universe are driven by a single matter component, which could be a barotropic perfect fluid or a scalar field (canonical or non-canonical). In that section, the effective speed of sound is introduced based on the requirement that the relations between quantities describing matter perturbations must be independent of the choice of gauge. This analysis is extended to a two-fluid/field-driven universe in Sec.~\ref{Sec: two fluid/field} and further generalized to a multi-fluid/field system in Sec.~\ref{Sec: multi fluid/field}. In Sec.~\ref{Sec: Curvature perturbation}, the evolution of curvature perturbations in a multi-fluid/field system is described, and it is shown that, at small scales, they satisfy a wave equation with the speed of sound defined in Sec.~\ref{Sec: multi fluid/field}. A closed set of equations governing the evolution of perturbations in a multi-fluid/field-driven universe is presented in Sec.~\ref{Sec: Closed set of equation}. Furthermore, in Sec.~\ref{Sec: Large scale behavior}, it is demonstrated that in a multi-fluid/field system, if perturbations are adiabatic at super-horizon scales, they remain so throughout the evolution of the universe. In Sec.~\ref{Sec: Dark Sector models}, the evolution of the effective speed of sound in unified dark sector models is discussed. Finally, in Sec.~\ref{Sec: conclusions}, the main results of this paper are summarized.

The conventions and notations used in this paper are as follows. All cosmological perturbations are described on a spatially flat Friedmann-Lema\^{\i}tre-Robertson-Walker (FLRW) metric with the signature $(+, -, -, -)$. The units are chosen such that the speed of light is set to unity ($c = 1$). An overdot denotes a derivative with respect to cosmic time, while a prime denotes a derivative with respect to conformal time.

\section{Equations of Cosmological Perturbations}\label{Sec: Basic eqn CPT}
In this section, we describe the equations governing the evolution of cosmological perturbations. We consider a spatially flat Friedmann-Lema\^{\i}tre-Robertson-Walker (FLRW) metric with scalar metric perturbations given by~\cite{Bardeen:1980kt,Kodama:1984ziu,Mukhanov:1990me}
\ber
ds^{2} = \l(1 + 2A\r)dt^{2} -  2a(t)B_{,i}dx^{i}dt - a^{2}(t)[(1 - 2 \psi)\delta_{ij}+ 2E_{,ij}]dx^{i}dx^{j},
  \label{Eqn: perturbed FRW line element}
\eer
where $a(t)$ is the scale factor, which depends only on time, while $A$, $\psi$, $E$, and $B$ are the scalar modes of the metric perturbation, which depend on both space and time. In Eq.~(\ref{Eqn: perturbed FRW line element}), $\delta_{ij}$ is the Kronecker delta function (where $i, j = 1, 2, 3$), and $B_{,i}$ denotes the partial derivative of $B$ with respect to the spatial coordinate $x^i$.

Under an infinitesimal coordinate transformation given by $x^{\mu} \rightarrow \tilde{x}^{\mu} = x^{\mu} + \xi^{\mu}$, where the $4$-vector $\xi^{\mu}$ describing the infinitesimal coordinate shift is decomposed as $\xi^{\mu} = (\xi^{0}, \delta^{i j}\xi_{,j})$, the variables describing metric perturbations in the line element (\ref{Eqn: perturbed FRW line element}) transform as~\cite{Sriramkumar:2012znv}
\ber
\widetilde{A} &=& A  - \dot{\xi}^{o},\label{Eqn: gauge transformation phi}\\
\widetilde{\psi} &=& \psi  + H\xi^{o},\label{Eqn: gauge transformation psi}\\
\widetilde{B} &=& B + a^{-1}\xi^{o} - a\dot{\xi},\label{Eqn: gauge transformation B}\\
\widetilde{E} &=& E - \xi,\label{Eqn: gauge transformation E}
\eer
In these equations, an overdot denotes a derivative with respect to time and $H = \dot{a}/a$. It is evident from the above equations that the variables describing the metric perturbations are gauge-dependent. This implies that even if we start with a homogeneous and isotropic FLRW metric at all scales, where ($A$, $\psi$, $E$, $B$) are zero in the line element (\ref{Eqn: perturbed FRW line element}), an infinitesimal coordinate transformation would result in non-zero values for ($\widetilde{A}$, $\widetilde{\psi}$, $\widetilde{E}$, $\widetilde{B}$). 
Such perturbations are not physical but are merely coordinate artifacts. To eliminate these types of artifacts from the real physical perturbations, one possibility is to fix the gauge by appropriately choosing $\xi^{0}$ and $\xi$, such as in the longitudinal gauge defined by $E = B = 0$~\cite{Mukhanov:1990me,Malik:2008im}. Another possibility is to describe the perturbations in terms of the gauge-invariant Bardeen potentials $\Phi$ and $\Psi$, defined as~\cite{Bardeen:1980kt}:
\ber
\Phi &=& A + \l[a(B - a\dot{E})\r]^{{\textbf{.}}},\label{Eqn: Bardeen Potential Phi}\\
\Psi &=& \psi - aH(B - a\dot{E})\label{Eqn: Bardeen Potential Psi}
\eer
It turns out that both approaches, namely, describing perturbations in the longitudinal gauge or in terms of gauge-invariant Bardeen potentials, are equivalent, as one obtains the same perturbation equations in both cases.

The energy-momentum tensor for both barotropic perfect fluids and scalar fields can be expressed as
\begin{equation}
T^{\mu}_{\hspace{0.2cm}\nu} = \left(\rho + p\right)u^{\mu}u_{\nu} - p\delta^{\mu}_{\hspace{0.2cm}\nu}, \label{Eqn: EM tensor}
\end{equation}
where $\rho$ is the energy density, $p$ is the pressure, and $u^{\mu}$ is the four-velocity field. We define the perturbations in the energy density $\rho$, the pressure $p$, and the four-velocity field $u^{\mu}$ as follows:
\begin{eqnarray}
\rho(t,\vec{x}) &=& \bar{\rho}(t) + \delta \rho(t,\vec{x}), \label{Eqn: density perturbation PF}\\
p(t,\vec{x}) &=&  \bar{p}(t) + \delta p(t,\vec{x}), \label{Eqn: pressure perturbation PF}\\
u^{\mu} &=& \bar{u}^{\mu} + \delta u^{\mu}, \label{Eqn: velocity perturbation PF}
\end{eqnarray}
where $\bar{\rho}(t)$, $\bar{p}(t)$, and $\bar{u}^{\mu} = [1,0,0,0]$ represent the energy density, pressure, and four-velocity field, respectively, in the background homogeneous and isotropic FLRW metric:
\beq
ds^{2} = dt^{2} - a^{2}(t)[dx^{2}+dy^{2}+dz^{2}]. \label{Eqn: FRW line element}
\eeq

Restricting ourselves to scalar metric perturbations as described by the line element (\ref{Eqn: perturbed FRW line element}), the perturbation in the four-velocity field can be decomposed as $\delta u^{\mu} = (\delta u^{0},\, \delta^{ij}u_{\hspace{0.05cm},\hspace{0.05cm}j})$. Since $u^{\mu}u_{\mu} = \bar{u}^{\mu}\bar{u}_{\mu} = 1$ and $u_{0} = g_{0\mu}u^{\mu}$, it turns out that $\delta u^{0} = - \delta u_{0} = -A$, where $A = \delta g_{00}/2$. Hence, the temporal part of $\delta u^{\mu}$ is not an independent quantity for the matter perturbation; rather, it depends solely on the metric perturbation $\delta g_{00}$.

At linear order in the perturbations, the energy-momentum tensor~(\ref{Eqn: EM tensor}) can be expressed as $T^{\mu}_{\hspace{0.2cm} \nu} = \bar{T}^{\mu}_{\hspace{0.2cm} \nu} + \delta T^{\mu}_{\hspace{0.2cm} \nu}$, where $\overline{T}^{\mu}_{\hspace{0.2cm} \nu}$ is the background energy-momentum tensor and $\delta T^{\mu}_{\hspace{0.2cm} \nu}$ represents the perturbations in the energy-momentum tensor. Using Eq.~(\ref{Eqn: EM tensor}) together with Eqs.~(\ref{Eqn: density perturbation PF}) through (\ref{Eqn: velocity perturbation PF}), it turns out that for the background $\overline{T}^{\mu}_{\hspace{0.2cm} \nu}$:
\ber
\overline{T}^{0}_{\hspace{0.2cm} 0} &=& \bar{\rho}(t)\\
\overline{T}^{0}_{\hspace{0.2cm} i} &=& 0\\
\overline{T}^{i}_{\hspace{0.2cm} j} &=& -\bar{p}(t)\delta^{i}_{\hspace{0.2cm}j}
\eer
whereas for the perturbed part $\delta T^{\mu}_{\hspace{0.2cm} \nu}$:
\ber
\delta T^{0}_{\hspace{0.2cm} 0} &=& \delta \rho\\
\delta T^{0}_{\hspace{0.2cm} i} &=& -\delta q_{,i}\label{Eqn: Defnition delta T 0i}\\
\delta T^{i}_{\hspace{0.2cm} j} &=& -\delta p \delta^{i}_{\hspace{0.2cm}j}
\eer
where
\beq
\delta q = a^{2}\l(\bar{\rho}+\bar{p}\r)\l[u + \frac{B}{a}\r]\label{Eqn: Definition of delta q}
\eeq
and the indices $i$ and $j$ take values from $1$ to $3$.

At the homogeneous and isotropic background level, the Einstein equations $\overline{G}^{\mu}_{\hspace{0.2cm} \nu} = (8 \pi G) \overline{T}^{\mu}_{\hspace{0.2cm} \nu}$ lead to the following Friedmann equations:
\ber
\frac{\dot{a}^2}{a^2} &=& \l(\frac{8 \pi G }{3}\r)\bar{\rho},\label{Eqn: Friedman Eqn-1}\\
\frac{\ddot{a}}{a} &=& -\l(\frac{4 \pi G }{3}\r)\l[\bar{\rho} + 3\bar{p}\r],\label{Eqn: Friedman Eqn-2}
\eer
whereas the linearized perturbed Einstein equations $\delta G^{\mu}_{\hspace{0.2cm} \nu} = (8 \pi G) \delta T^{\mu}_{\hspace{0.2cm} \nu}$ imply that:
\ber
3H^{2}A + 3H\dot{\psi} + \frac{k^{2}}{a^{2}}\left[\psi - aH\left(aB - a^{2}\dot{E}\right)\right] 
 &=& - 4 \pi G \delta \rho,\label{Eqn: Perturbed Einstein Eqn-1}\\
\dot{\psi} + HA  &=& - 4 \pi G \delta q,\label{Eqn: Perturbed Einstein Eqn-2}\\
\ddot{\psi} + 3H\dot{\psi} + H\dot{A} + \left(2\dot{H} + 3H^{2}\right)A &=& 4 \pi G \delta p,\label{Eqn: Perturbed Einstein Eqn-3}\\
A - \psi + H\left(aB - a^{2}\dot{E}\right) + \left(aB - a^{2}\dot{E}\right)^{\textbf{.}} &=&  0.\label{Eqn: no anisotropic stress Eqn}
\eer
In Eq.~(\ref{Eqn: Perturbed Einstein Eqn-1}), $k = 2 \pi /\lambda$ is the wavenumber, which is inversely proportional to the length scale of the perturbation, $\lambda$. Furthermore, it is important to note that Eq.~(\ref{Eqn: no anisotropic stress Eqn}) is a consequence of the fact that the energy-momentum tensor~(\ref{Eqn: EM tensor}) has no anisotropic stress, \textit{i.e.}, $\delta T^{i}_{\hspace{0.2cm}j} = 0$ for all $i \neq j$ (where $i,j = 1,2,3$). From Eqs.~(\ref{Eqn: Bardeen Potential Phi}) and (\ref{Eqn: Bardeen Potential Psi}), this further implies that:
\beq
\Phi = \Psi \label{Eqn: equality of Bardeen Potentials}
\eeq

The covariant conservation equation, \textit{i.e.}, $T^{\mu}_{\hspace{0.2cm}\nu\, ;\, \mu} = 0$, implies that for the background variables $\bar{\rho}(t)$ and $\bar{p}(t)$:
\beq
\dot{\bar{\rho}} = -3H(\bar{\rho} + \bar{p})\label{Eqn: background conservation eqn}
\eeq
For the variables describing the matter perturbations, \textit{viz.}, $\delta \rho$, $\delta q$, and $\delta p$, one obtains the following two equations:
\ber
\dot{\delta\rho} + 3H(\delta \rho + \delta p) - 3(\bar{\rho} +\bar{ p})\dot{\psi} &=& \frac{k^{2}}{a^2}\l[\delta q - a(\bar{\rho} +\bar{ p})(B - a \dot{E})\r]\label{Eqn: delta rho conservation eqn}\\
\dot{\delta q} + 3H\delta q + \delta p + A(\bar{\rho} +\bar{ p}) &=& 0  \label{Eqn: delta q canservation eqn}
\eer

Gauge transformations for the variables describing matter perturbations $\delta \rho$, $\delta p$, and $\delta q$, resulting from an infinitesimal coordinate transformation, \emph{viz.}, $x^{\mu} \rightarrow \tilde{x}^{\mu} = x^{\mu} + \xi^{\mu}$, where $\xi^{\mu} = (\xi^{0}, \delta^{i j}\xi_{,j})$, are given by:
\ber
\widetilde{\delta \rho} &=& \delta \rho - \dot{\bar{\rho}}\xi^{o},\label{Eqn: gauge transformation delta rho}\\
\widetilde{\delta p} &=& \delta p - \dot{\bar{p}}\xi^{o},\label{Eqn: gauge transformation delta p}\\
\widetilde{\delta q} &=& \delta q + (\bar{\rho} +\bar{ p})\xi^{o}.\label{Eqn: gauge transformation delta q}
\eer
Therefore, the variables describing matter perturbations, \emph{i.e.}, $\delta \rho$, $\delta p$, and $\delta q$, are gauge-dependent. However, just like the gauge-invariant Bardeen potentials $\Phi$ and $\Psi$, one can construct the following gauge-invariant matter variables $\delta \rho^{^{(gi)}}_{_B}$, $\delta p^{^{(gi)}}_{_B}$, and $\delta q^{^{(gi)}}_{_B}$:
\ber
\delta \rho^{^{(gi)}}_{_B} &=& \delta \rho + a \dot{\bar{\rho}}(B - a\dot{E}),\label{Eqn: gauge invarian delta rho}\\
\delta p^{^{(gi)}}_{_B} &=& \delta p + a \dot{\bar{p}}(B - a\dot{E}),\label{Eqn: gauge invarian delta p}\\
\delta q^{^{(gi)}}_{_B} &=& \delta q - a(\bar{\rho} +\bar{ p})(B - a \dot{E}).\label{Eqn: gauge invarian delta q}
\eer

The Eqs.~(\ref{Eqn: Perturbed Einstein Eqn-1}) through (\ref{Eqn: Perturbed Einstein Eqn-3}), which describe the perturbed Einstein equations $\delta G^{\mu}_{\hspace{0.2cm} \nu} = (8 \pi G) \delta T^{\mu}_{\hspace{0.2cm} \nu}$, can now be expressed in terms of the gauge-invariant Bardeen potential $\Phi$ using Eqs.~(\ref{Eqn: Bardeen Potential Phi}), (\ref{Eqn: Bardeen Potential Psi}), and (\ref{Eqn: equality of Bardeen Potentials}) together with Eqs.~(\ref{Eqn: gauge invarian delta rho}) through (\ref{Eqn: gauge invarian delta q}). The gauge-invariant perturbed Einstein equations thus obtained are given by:
\ber
3H^{2}\Phi + 3H\dot{\Phi} + \frac{k^{2}}{a^{2}}\Phi = - (4 \pi G)\delta \rho^{^{(gi)}}_{_B}, \label{Eqn: GI Perturbed Einstein Eqn-1}\\
\dot{\Phi} + H\Phi  = - (4 \pi G) \delta q^{^{(gi)}}_{_B}, \label{Eqn: GI Perturbed Einstein Eqn-2}\\
\ddot{\Phi} + 4H\dot{\Phi} + \left(2\dot{H} + 3H^{2}\right)\Phi = (4 \pi G) \delta p^{^{(gi)}}_{_B} \label{Eqn: GI Perturbed Einstein Eqn-3}
\eer
Similarly, we can express the conservation equations for matter perturbations, given by Eqs.~(\ref{Eqn: delta rho conservation eqn}) and (\ref{Eqn: delta q canservation eqn}), in terms of the gauge-invariant quantities $\delta \rho^{^{(gi)}}_{_B}$, $\delta q^{^{(gi)}}_{_B}$, and $\Phi$. The conservation equations thus obtained are:
\ber
\dot{\delta \rho}^{^{(gi)}}_{_B} + 3H(\delta \rho^{^{(gi)}}_{_B} + \delta p^{^{(gi)}}_{_B})- 3(\bar{\rho} +\bar{p})\dot{\Phi} = \l(\frac{k^{2}}{a^2}\r)\delta q^{^{(gi)}}_{_B}\label{Eqn: GI delta rho conservation eqn}\\
\dot{\delta q}^{^{(gi)}}_{_B}+ 3H\delta q^{^{(gi)}}_{_B}+\delta p^{^{(gi)}}_{_B}+(\bar{\rho} +\bar{p})\Phi =  0 \label{Eqn: GI delta q conservation eqn}
\eer
The above equations, together with Eqs.~(\ref{Eqn: GI Perturbed Einstein Eqn-1}) through (\ref{Eqn: GI Perturbed Einstein Eqn-3}), represent the gauge-invariant equations governing the evolution of cosmological perturbations. These equations are also identical to those obtained in the longitudinal gauge, defined as the gauge in which $E = B = 0$.

\section{The case of single fluid/field cosmological system}\label{Sec: single fluid/field}
Let us first consider the case where the dynamics of the universe are driven either by a single barotropic perfect fluid or by a single canonical or non-canonical scalar field. For both types of matter, the energy-momentum tensor can be expressed in the form of a perfect fluid, as given in Eq.~(\ref{Eqn: EM tensor}). By definition, for a barotropic perfect fluid, the pressure is a function of energy density only, \textit{i.e.}, $p = p(\rho)$, at both the background and perturbation levels. However, this relation does not generally hold for scalar fields.

The evolution of the scale factor $a(t)$ is determined by the Friedmann equation~(\ref{Eqn: Friedman Eqn-1}) and the background conservation equation~(\ref{Eqn: background conservation eqn}). However, to solve Eq.~(\ref{Eqn: background conservation eqn}), a relation between the background pressure $\bar{p}$ and the energy density $\bar{\rho}$ is required. The equation of state parameter $w$ is defined as the ratio of the background pressure to the background energy density. For barotropic perfect fluids, $w$ is a constant parameter characterizing the fluid; for example, $w = 0$ for cold dark matter and $w = 1/3$ for radiation. In the case of scalar fields, the parameter $w$ is determined by the field equations and, in general, may be a function of time. Once the relation $w(t) = \bar{p}(t)/\bar{\rho}(t)$ is specified, the evolution of the scale factor $a(t)$ is fully determined by the closed set of equations~(\ref{Eqn: Friedman Eqn-1}) and (\ref{Eqn: background conservation eqn}).

In a similar way, to determine the evolution of perturbations using Eqs.~(\ref{Eqn: Perturbed Einstein Eqn-1}) through (\ref{Eqn: Perturbed Einstein Eqn-3}) alongside Eqs.~(\ref{Eqn: delta rho conservation eqn}) and (\ref{Eqn: delta q canservation eqn}), or equivalently using Eqs.~(\ref{Eqn: GI Perturbed Einstein Eqn-1}) through (\ref{Eqn: GI delta q conservation eqn}), it is necessary to specify how $\delta p$ is related to $\delta \rho$ and $\delta q$. At linear order in the perturbations, we expect a linear relation between these variables. Therefore, we can generally express $\delta p$ as:
\beq
\delta p = c_{1}\delta \rho +  c_{2}\delta q,\label{Eqn: relation delta rho p and q}
\eeq
where $c_{1}$ and $c_{2}$ are background-dependent quantities that may be functions of time. 
It is important to note that Eq.~(\ref{Eqn: relation delta rho p and q}) is valid when there is one scalar degree of freedom obeying a second-order differential equation, such that initial conditions are specified by two functions of space and time (e.g., the density and velocity) with no additional (internal) degrees of freedom. This is true for both barotropic perfect fluids and scalar fields (canonical and non-canonical scalar fields). We shall see in Eq.~(\ref{Eqn: SF Phi dot dot Eqn}) that the evolution of cosmological perturbations in such a system obeying Eq.~(\ref{Eqn: relation delta rho p and q}) can be determined from a second-order differential equation satisfied by the Bardeen potential $\Phi$.

The relation~(\ref{Eqn: relation delta rho p and q}) must be a consequence of the intrinsic properties of the matter content; that is, it must reflect the nature of the single barotropic perfect fluid or scalar field driving the dynamics of the universe. Consequently, this relation must be gauge-invariant in form. This implies that after an infinitesimal coordinate transformation $x^{\mu} \rightarrow \tilde{x}^{\mu} =  x^{\mu} + \xi^{\mu}$ (where $\xi^{\mu} = (\xi^{0}, \delta^{i j}\xi_{,j})$), the transformed variables $\widetilde{\delta p}$, $\widetilde{\delta \rho}$, and $\widetilde{\delta q}$ must satisfy the same functional form as Eq.~(\ref{Eqn: relation delta rho p and q}):
\beq
\widetilde{\delta p} = c_{1}\widetilde{\delta \rho} +  c_{2}\widetilde{\delta q},\label{Eqn: relation delta rho p and q new gauge}
\eeq
where $c_{1}$ and $c_{2}$ are the same parameters that appear in Eq.~(\ref{Eqn: relation delta rho p and q}). For Eqs.~(\ref{Eqn: relation delta rho p and q}) and (\ref{Eqn: relation delta rho p and q new gauge}) to be consistent with the gauge transformations (\ref{Eqn: gauge transformation delta rho}) through (\ref{Eqn: gauge transformation delta q}), we require:
\beq
c_{2} = 3H(c_{a}^2 - c_{1}),\label{Eqn: c1 c2}
\eeq
where
\beq
c_{a}^2 = \frac{\dot{\bar{p}}}{\dot{\bar{\rho}}}\label{Eqn: ca2 adiabatic speed of sound}
\eeq
is defined as the square of the adiabatic speed of sound. The rationale for defining $c_{a}$ as the adiabatic speed of sound is that, as will be demonstrated later, in the absence of non-adiabatic pressure perturbations, $c_{a}$ is indeed the speed at which the gauge-invariant Bardeen potential propagates at small scales.

The comoving gauge corresponds to a gauge in which $u = B = 0$, and consequently, from Eq.~(\ref{Eqn: Definition of delta q}), it follows that $\delta q = 0$. Therefore, in this gauge, the off-diagonal terms in the perturbed energy-momentum tensor vanish, \textit{i.e.}, $\delta T^{\mu}_{\hspace{0.2cm}\nu} = 0$ for all $\mu \neq \nu$ (where $\mu,\nu = 0, 1, 2, 3$), and the matter perturbations are defined solely by $\delta p$ and $\delta \rho$. Defining the square of the effective speed of sound, $c_{e}^{2}$, for a single fluid/field system as the ratio $\delta p/\delta \rho$ in the comoving gauge ($\delta q = 0$), we find from Eq.~(\ref{Eqn: relation delta rho p and q}) that $c_{1} = c_{e}^{2}$. Therefore, using this relation and Eq.~(\ref{Eqn: c1 c2}), we can express Eq.~(\ref{Eqn: relation delta rho p and q}) as follows:
\beq
\delta p = c_{e}^{2}\,\delta \rho +  3H(c_{a}^{2} - c_{e}^2) \delta q.\label{Eqn: SF relation delta rho p and q}
\eeq
The above equation~(\ref{Eqn: SF relation delta rho p and q}) is valid when the dynamics of the universe are driven by a single matter component, which can be either a barotropic perfect fluid or a canonical or non-canonical scalar field.

For a barotropic perfect fluid, since the pressure perturbation is directly proportional to the energy density perturbation in every gauge, Eq.~(\ref{Eqn: SF relation delta rho p and q}) implies that $c_{e}^{2} = c_{a}^2$ for such fluids. We treat this as the defining characteristic of such a fluid; \textit{i.e.}, we define a barotropic perfect fluid as one for which the effective speed of sound $c_{e}$ is identically equal to the adiabatic speed of sound $c_{a}$.

Let us now consider the case where the dynamics of the universe are driven by a single scalar field, either canonical or non-canonical. The Lagrangian of such a general scalar field is given by:
\beq
\mathcal{L} = \mathcal{L}(X,\phi),\label{Eqn: Lagrangian NC scalar field}
\eeq
where
\beq
X = \f{1}{2}\partial_\mu\phi\partial^\mu\phi
\eeq
is the kinetic term of the scalar field $\phi$.

The energy-momentum tensor of the scalar field $\phi$ corresponding to the Lagrangian~(\ref{Eqn: Lagrangian NC scalar field}) is given by:
\beq
T^{\mu}_{\hspace{0.2cm}\nu} = \frac{\partial\mathcal{L}}{\partial X}\partial^{\mu}\phi\partial_{\nu}\phi - \mathcal{L}\delta^{\mu}_{\hspace{0.2cm}\nu}
\label{Eqn: Scalar field EM tensor}.
\eeq
The perturbation in the scalar field is defined as:
\beq
\phi(t, \vec{x}) = \bar{\phi}(t) + \delta \phi(t, \vec{x})\label{Eqn: perturbation in filed phi}
\eeq
where $\bar{\phi}(t)$ is the background field which governs the evolution of the scale factor. Under an infinitesimal coordinate transformation $x^{\mu} \rightarrow \tilde{x}^{\mu} =  x^{\mu} + \xi^{\mu}$, where $\xi^{\mu} = (\xi^{0}, \delta^{i j}\xi_{,j})$, the scalar field perturbation transforms as:
\beq
\widetilde{\delta \phi} = \delta \phi - \dot{\bar{\phi}}\xi^{0}
\eeq
Using Eq.~(\ref{Eqn: perturbation in filed phi}), we can express the energy-momentum tensor~(\ref{Eqn: Scalar field EM tensor}) in the same form as that of the perfect fluid given in Eq.~(\ref{Eqn: EM tensor}). Consequently, it turns out that:
\ber
\bar{\rho} &=& 2X\mathcal{L}_{X} - \mathcal{L},\label{Eqn: density non canonical scalar field}\\
\bar{p}&=& \mathcal{L} \label{Eqn: pressure non canonical scalar field}
\eer
where $\bar{\rho}$ and $\bar{p}$ correspond to the energy density and pressure of the scalar field, respectively. In Eqs.~(\ref{Eqn: density non canonical scalar field}) and (\ref{Eqn: pressure non canonical scalar field}), we have considered $\phi = \bar{\phi}(t)$, such that $X = (1/2)\dot{\bar{\phi}}^2$. Note that $\mathcal{L}_{X}$ in Eq.~(\ref{Eqn: density non canonical scalar field}) is defined as the partial derivative of $\mathcal{L}$ with respect to $X$.

Similarly, by substituting Eq.~(\ref{Eqn: perturbation in filed phi}) into Eq.~(\ref{Eqn: Scalar field EM tensor}) and using Eq.~(\ref{Eqn: EM tensor}), we find that for the scalar field, $\delta \rho$, $\delta p$, and $\delta q$ are given by:
\ber
\delta\rho &=& \l[\mathcal{L}_{X} + 2X \mathcal{L}_{XX}\r]\l(\dot{\bar{\phi}}\dot{\delta\phi} - A\dot{\bar{\phi}}^2\r) -  \l[\mathcal{L}_{\phi} - 2X \mathcal{L}_{X\phi}\r]\delta\phi,
\label{Eqn: perturbed density NC scalar field}\\
\delta p &=& \mathcal{L}_{X}\l(\dot{\bar{\phi}}\dot{\delta\phi} - A\dot{\bar{\phi}}^2\r)+  \mathcal{L}_{\phi}\delta\phi,
\label{Eqn: perturbed pressure NC scalar field}\\
\delta q &=& -\dot{\bar{\phi}}\mathcal{L}_{X}\delta \phi. \label{Eqn: perturbed delta q NC scalar field}
\eer
Note that in the above three equations, $\mathcal{L}_{X}$, $\mathcal{L}_{XX}$, $\mathcal{L}_{\phi}$, and $\mathcal{L}_{X\phi}$ are all background-dependent quantities. By substituting Eqs.~(\ref{Eqn: perturbed density NC scalar field}) through (\ref{Eqn: perturbed delta q NC scalar field}) into Eq.~(\ref{Eqn: SF relation delta rho p and q}), we obtain:
\beq
c_{e}^{2} = \frac{\mathcal{L}_{X}}{\mathcal{L}_{X} + 2X \mathcal{L}_{XX}}.\label{Eqn: c_e^2 for NC scalar field}
\eeq
This is the square of the effective speed of sound for a general minimally coupled scalar field with Lagrangian $\mathcal{L}(X,\phi)$, as introduced by Garriga and Mukhanov in Ref.~\cite{Garriga:1999vw}. The effective speed of sound thus defined is a background-dependent, space-independent quantity and is therefore gauge-invariant; by contrast, the ratio of the pressure perturbation to the energy density perturbation is, in general, a gauge-dependent quantity.

For canonical scalar fields, since $\mathcal{L}(X,\phi) = X - V(\phi)$, the effective speed of sound turns out to be exactly equal to unity, irrespective of the form of the potential $V(\phi)$~\cite{Hu:1998kj}. This is evident from Eq.~(\ref{Eqn: c_e^2 for NC scalar field}). However, for non-canonical scalar fields, $c_e^2$ can take any value less than unity, depending on the form of the Lagrangian. For example, in models where $\mathcal{L}(X,\phi) = X^{\alpha} - V(\phi)$ and $\alpha$ is a constant, it follows from Eq.~(\ref{Eqn: c_e^2 for NC scalar field}) that $c_{e}^{2} = (2\alpha -1)^{-1}$~\cite{Unnikrishnan:2012zu,Li:2012vta,Fang:2006yh,Mukhanov:2005bu}. It is important to note that the effective speed of sound for a scalar field is not, in general, equal to the adiabatic speed of sound, unlike the case of barotropic perfect fluids for which $c_e^2 = c_a^2$. For instance, in the case of canonical scalar fields $c_e^2 = 1$, whereas the value of $c_a^2 = \dot{\bar{p}}/\dot{\bar{\rho}}$ depends on the form of the potential $V(\phi)$ in the Lagrangian $\mathcal{L}(X,\phi) = X - V(\phi)$.

In general, we can express the pressure perturbation as:
\beq
\delta p = c_{a}^2\,\delta \rho + \delta p_{\mathrm{nad}}\label{Eqn: delta p nad intro}
\eeq
where $\delta p_{\mathrm{nad}}$ represents the non-adiabatic pressure perturbations. Note that if $\delta p_{\mathrm{nad}} = 0$, then $\delta p = c_{a}^2\,\delta \rho$ in every gauge of cosmological perturbations; \textit{i.e.}, even in a new gauge defined by Eqs.~(\ref{Eqn: gauge transformation delta rho}) and (\ref{Eqn: gauge transformation delta p}), one finds that $\widetilde{\delta p} = c_{a}^2\,\widetilde{\delta \rho}$. It is for this reason that $c_{a}$ is called the adiabatic speed of sound. It is with this speed that perturbations propagate at small scales when they are adiabatic, \textit{i.e.}, when $\delta p_{\mathrm{nad}} = 0$. Using Eq.~(\ref{Eqn: SF relation delta rho p and q}), we find that:
\beq
\delta p_{\mathrm{nad}} = (c_{e}^{2}- c_{a}^2)\l[\delta \rho - 3H \delta q\r]\label{Eqn: delta p nad eqn single field}
\eeq
This implies that the difference between the two speeds of sound results in a non-zero non-adiabatic pressure perturbation in scalar field models for which, in general, $c_{e}^{2} \ne c_{a}^2$. Note that for barotropic perfect fluids, $\delta p_{\mathrm{nad}} = 0$ since $c_{e}^{2} = c_{a}^2$. Similarly, for pure kinetic non-canonical scalar field models, it turns out that $c_{e}^{2} = c_{a}^2$, and therefore $\delta p_{\mathrm{nad}} = 0$. Consequently, such scalar fields behave cosmologically as barotropic perfect fluids~\cite{Unnikrishnan:2010ag, Arroja:2010wy}.

Note that by using Eqs.~(\ref{Eqn: GI Perturbed Einstein Eqn-1}) and (\ref{Eqn: GI Perturbed Einstein Eqn-2}), Eq.~(\ref{Eqn: delta p nad eqn single field}) can be expressed as:
\beq
\delta p_{\mathrm{nad}} = \l(\frac{c_{a}^2 - c_{e}^2}{4 \pi G a^2}\r)k^2\Phi\label{Eqn: delta p nad eqn single field k dependence}
\eeq
Evidently, $\delta p_{\mathrm{nad}} \rightarrow 0$ as $k \rightarrow 0$ when the dynamics of the universe are driven by a single matter component, which could be either a barotropic perfect fluid or a scalar field (canonical or non-canonical)~\cite{Unnikrishnan:2010ag,Christopherson:2008ry}.

The gauge-invariant variables $\delta \rho^{^{(gi)}}_{_B}$, $\delta p^{^{(gi)}}_{_B}$, and $\delta q^{^{(gi)}}_{_B}$, defined in Eqs.~(\ref{Eqn: gauge invarian delta rho}), (\ref{Eqn: gauge invarian delta p}), and (\ref{Eqn: gauge invarian delta q}), respectively, are related in exactly the same way that $\delta p$ is related to $\delta \rho$ and $\delta q$ in Eq.~(\ref{Eqn: SF relation delta rho p and q}). Therefore, we can combine Eqs.~(\ref{Eqn: GI Perturbed Einstein Eqn-1}) through (\ref{Eqn: GI Perturbed Einstein Eqn-3}) to obtain the following evolution equation for the gauge-invariant Bardeen potential $\Phi$:
\ber
\ddot{\Phi} + H\l[4 + 3c_a^2\r]\dot{\Phi} + \l[2\dot{H} + 3H^{2}(1 + c_a^2)\r]\Phi + \l(\f{c_e^2 k^2}{a^2}\r)\Phi \,=\,0.\label{Eqn: SF Phi dot dot Eqn}
\eer

On scales much smaller than the Hubble radius, \textit{i.e.}, in the large-$k$ limit where $c_e k \gg aH$, Eq.~(\ref{Eqn: SF Phi dot dot Eqn}) implies that:
\beq
\ddot{\Phi}\, \simeq \, - \l(\f{c_e^2 k^2}{a^2}\r)\Phi \label{Eqn: wave eqn Phi}
\eeq
This is a wave equation where $c_e$ represents the speed of sound. Consequently, the effective speed of sound is the speed at which the gauge-invariant Bardeen potential propagates on scales much smaller than the Hubble radius. Recall that $c_e$ for a general scalar field is given by Eq.~(\ref{Eqn: c_e^2 for NC scalar field}), whereas for a barotropic perfect fluid, it is $c_a = \sqrt{\dot{\bar{p}}/\dot{\bar{\rho}}}$. It is evident from Eq.~(\ref{Eqn: SF Phi dot dot Eqn}) that the effective speed of sound $c_e$ is a critical parameter determining the evolution of perturbations. Once the evolution of $\Phi$ is determined from Eq.~(\ref{Eqn: SF Phi dot dot Eqn}), the evolution of $\delta \rho$ and $\delta q$ in the longitudinal gauge can be obtained from Eqs.~(\ref{Eqn: GI Perturbed Einstein Eqn-1}) and (\ref{Eqn: GI Perturbed Einstein Eqn-2}), respectively, and $\delta p$ from Eq.~(\ref{Eqn: SF relation delta rho p and q}).

Just like the gauge-invariant Bardeen variable $\delta \rho^{^{(gi)}}_{_B}$ defined in Eq.~(\ref{Eqn: gauge invarian delta rho}), we can combine Eqs.~(\ref{Eqn: gauge transformation delta rho}) and (\ref{Eqn: gauge transformation delta q}) to define the following gauge-invariant density perturbation $\delta \rho^{^{(gi)}}_{_c}$:
\beq
\delta \rho^{^{(gi)}}_{_c}  = \delta \rho - 3H \delta q\label{Eqn: GI rho in comoing gauge}
\eeq
Similarly, the gauge-invariant pressure perturbation $\delta p^{^{(gi)}}_{_c}$ is defined as:
\beq
\delta p^{^{(gi)}}_{_c}  = \delta p - 3H c_{a}^2\delta q\label{Eqn: GI p in comoing gauge}
\eeq
These gauge-invariant quantities, $\delta \rho^{^{(gi)}}_{_c}$ and $\delta p^{^{(gi)}}_{_c}$, coincide with $\delta \rho$ and $\delta p$, respectively, in the comoving gauge\footnote{It is for this reason, and to differentiate it from the gauge-invariant Bardeen variable $\delta \rho^{^{(gi)}}_{_B}$, that we have used the subscript `$c$' in $\delta \rho^{^{(gi)}}_{_c}$.}, \textit{i.e.}, the gauge in which $\delta q = 0$. Note that it follows from Eq.~(\ref{Eqn: SF relation delta rho p and q}) that $c_e^2 = \delta p^{^{(gi)}}_{_c}/\delta \rho^{^{(gi)}}_{_c}$. Therefore, the effective speed of sound defined in Eq.~(\ref{Eqn: SF relation delta rho p and q}) is the ratio of these specific gauge-invariant pressure and energy density perturbations. While the ratio $\delta p/\delta \rho$ is generally a gauge-dependent quantity, this particular combination ensures that $c_e^2$ is not only gauge-invariant but also strictly background-dependent~\cite{Unnikrishnan:2008ki}.

Using Eqs.~(\ref{Eqn: GI Perturbed Einstein Eqn-1}) and (\ref{Eqn: GI Perturbed Einstein Eqn-2}), it follows that the gauge-invariant density perturbation $\delta \rho^{^{(gi)}}_{_c}$ satisfies the following Poisson equation:
\beq
k^2 \Phi = -(4\pi G a^ 2)\,\delta \rho^{^{(gi)}}_{_c}.\label{Eqn: Poisson eqn for Phi}
\eeq
Substituting Eq.~(\ref{Eqn: Poisson eqn for Phi}) into Eq.~(\ref{Eqn: SF Phi dot dot Eqn}), we obtain the evolution equation for $\delta \rho^{^{(gi)}}_{_c}$:
\ber
\ddot{\delta \rho}^{^{(gi)}}_{_c}  + H\l[8 + 3c_a^2 \r]\dot{\delta \rho}^{^{(gi)}}_{_c}\, + \l[4\dot{H}+ 3H^{2}(5 + 3c_a^2)\r]\delta \rho^{^{(gi)}}_{_c} + \l(\f{c_e^2 k^2}{a^2}\r)\delta \rho^{^{(gi)}}_{_c}\, =\, 0. \label{Eqn: SF comiving delta rho dot dot Eqn}
\eer
Consequently, analogous to Eq.~(\ref{Eqn: wave eqn Phi}), on scales much smaller than the Hubble radius, \textit{i.e.}, in the large-$k$ limit, $\delta \rho^{^{(gi)}}_{_c}$ also satisfies the same wave equation, \textit{viz.}:
\beq
\ddot{\delta \rho}^{^{(gi)}}_{_c}    \, \simeq \, - \l(\f{c_e^2 k^2}{a^2}\r)\delta \rho^{^{(gi)}}_{_c}. \label{Eqn: wave eqn delta rho}
\eeq
Therefore, the gauge-invariant $\delta \rho^{^{(gi)}}_{_c}$, or equivalently the density perturbation $\delta \rho$ in the comoving gauge ($\delta q = 0$), propagates with the effective speed of sound $c_e$ on scales much smaller than the Hubble radius.

\section{\label{Sec: two fluid/field}The case of two fluid/field cosmological system}

Let us now consider the case where the dynamics of the universe are driven by two minimally coupled matter components. These may both be non-canonical scalar fields, barotropic perfect fluids, or a combination of a non-canonical scalar field and a barotropic perfect fluid. Note that by non-canonical scalars, we mean scalar fields whose Lagrangians are general functions of the kinetic term $X = (1/2)\partial_\mu\phi\partial^\mu\phi$ and the field $\phi$, as expressed in Eq.~(\ref{Eqn: Lagrangian NC scalar field}). This formulation also encompasses the case of a canonical scalar field, for which $\mathcal{L}(X,\phi) = X - V(\phi)$.

As described in the preceding section, the energy-momentum tensors of both barotropic perfect fluids and non-canonical scalar fields can be expressed in the form of a perfect fluid, as given in Eq.~(\ref{Eqn: EM tensor}). The total energy-momentum tensor can be expressed as the sum of the individual components:
\beqstr
T^{\mu}_{\hspace{0.2cm}\nu}\,=\, ^{^{(1)}}T^{\mu}_{\hspace{0.2cm}\nu}\,+\,^{^{(2)}}T^{\mu}_{\hspace{0.2cm}\nu},\nonumber
\eeqstr
where $^{^{(1)}}T^{\mu}_{\hspace{0.2cm}\nu}$ and $^{^{(2)}}T^{\mu}_{\hspace{0.2cm}\nu}$ are the energy-momentum tensors of the two matter components, respectively. Therefore, at the background level, we have:
\berstr
\bar{\rho}\,&=&\,\bar{\rho}_{_1} + \bar{\rho}_{_2},\\
\bar{p}\,&=&\,\bar{p}_{_1} + \bar{p}_{_2}
\eerstr

The evolution of the scale factor is governed by the Friedmann equation~(\ref{Eqn: Friedman Eqn-1}) along with the conservation equations for each matter component, as given by:
\ber
\frac{\dot{a}^2}{a^2} &=& \l(\frac{8 \pi G }{3}\r)\l[\bar{\rho}_{_1} + \bar{\rho}_{_2}\r],\label{Eqn: Friedman Eqn-1 two flied/fluid case}\\
\dot{\bar{\rho}}_{_1} &=& -3H\l(\bar{\rho}_{_1} + \bar{p}_{_1}\r)\label{Eqn: background conservation eqn matter content 1}\\
\dot{\bar{\rho}}_{_2} &=& -3H\l(\bar{\rho}_{_2} + \bar{p}_{_2}\r)\label{Eqn: background conservation eqn matter content 2}
\eer
We assume the two matter components have no direct interaction beyond gravity, and therefore Eqs.~(\ref{Eqn: background conservation eqn matter content 1}) and (\ref{Eqn: background conservation eqn matter content 2}) represent conservation equations for each component separately.

If $\omega_{_1}$ and $\omega_{_2}$ are the equation of state parameters for each matter component, respectively, then the equation of state parameter $\omega$ of the total system is given by:
\beq
\omega = \frac{\omega_{_1}\,\bar{\rho}_{_1}\,+\, \omega_{_2}\,\bar{\rho}_{_2}}{\bar{\rho}_{_1} + \bar{\rho}_{_2}}\label{Eqn: Total EOS two flied/fluid case}
\eeq

At first order in the perturbations, since $\delta T^{\mu}_{\hspace{0.2cm}\nu}\,=\, ^{^{(1)}}\delta T^{\mu}_{\hspace{0.2cm}\nu}\,+\,^{^{(2)}}\delta T^{\mu}_{\hspace{0.2cm}\nu}$, we obtain from Eq.~(\ref{Eqn: EM tensor}):
\ber
\delta \rho \,&=&\,\delta \rho_{_1} +\delta \rho_{_2},\label{Eqn: delta rho two field/fluid}\\
\delta p\,&=&\,\delta p_{_1} + \delta p_{_2},\label{Eqn: delta p two field/fluid}\\
\delta q \,&=&\,\delta q_{_1} +  \delta q_{_2},\label{Eqn: delta q two field/fluid}
\eer
where $\delta q$ is defined in Eq.~(\ref{Eqn: Definition of delta q}). Note that in Eqs.~(\ref{Eqn: delta rho two field/fluid}) through (\ref{Eqn: delta q two field/fluid}), the quantities $\delta \rho$, $\delta p$, and $\delta q$ define the perturbations in the matter content of the total system.

For both barotropic perfect fluids and non-canonical scalars, the relationship between the pressure perturbation $\delta p$, the energy density perturbation $\delta \rho$, and $\delta q$ is given by Eq.~(\ref{Eqn: SF relation delta rho p and q}). Therefore, for both matter components, we can express the pressure perturbations as:
\ber
\delta p_{_1} &=& c_{e1}^{2}\,\delta \rho_{_1} +  3H\l(c_{a1}^{2} - c_{e1}^2\r) \delta q_{_1},\label{Eqn: TwoF relation delta rho p1 and q1}\\
\delta p_{_2} &=& c_{e2}^{2}\,\delta \rho_{_2} +  3H\l(c_{a2}^{2} - c_{e2}^2\r) \delta q_{_2},\label{Eqn: TwoF relation delta rho p2 and q2}
\eer
where $c_{e1}$ and $c_{e2}$ are the effective speeds of sound for the two matter components, respectively, and
\berstr
c_{a1}^{2} &=& \frac{\dot{\bar{p}}_{_1}}{\dot{\bar{\rho}}_{_1}},\\
c_{a2}^{2} &=& \frac{\dot{\bar{p}}_{_2}}{\dot{\bar{\rho}}_{_2}},
\eerstr
are the squares of the adiabatic speeds of sound for the two components. Recall that for barotropic perfect fluids, the effective speed of sound coincides with the adiabatic speed of sound. However, for non-canonical scalars, the effective speed of sound is given by Eq.~(\ref{Eqn: c_e^2 for NC scalar field}), which in general may not be equal to the adiabatic speed of sound.

The primary aim of this paper is to determine how the total pressure perturbation $\delta p$ is related to $\delta \rho$ and $\delta q$, thereby defining the effective speed of sound for the total system. At first order in cosmological perturbation theory, we expect $\delta p$ to be linearly related to $\delta \rho$ and $\delta q$, analogous to Eq.~(\ref{Eqn: relation delta rho p and q}). For such a relation to be consistent across every gauge, we should obtain an equation similar in form to Eq.~(\ref{Eqn: SF relation delta rho p and q}). However, since the dynamics of the universe are driven by two matter components, we expect the following two gauge-invariant entropy perturbations (or isocurvature perturbations) to contribute to the evolution of the system:
\beqstr
\l[\frac{\delta \rho_{_1}}{\bar{\rho}_{_1} + \bar{p}_{_1}}\,-\,\frac{\delta \rho_{_2}}{\bar{\rho}_{_2} + \bar{p}_{_2}}\r]~~~~\mathrm{and}~~~~\l[\frac{\delta q_{_1}}{\bar{\rho}_{_1} + \bar{p}_{_1}}\,-\,\frac{\delta q_{_2}}{\bar{\rho}_{_2} + \bar{p}_{_2}}\r].\nonumber
\eeqstr
The rationale for this is as follows. Isocurvature (or entropy) perturbations correspond to those modes that represent a deviation from adiabaticity~\cite{Lyth:2009imm,Gordon:2000hv,Bartolo:2001rt}. These modes result from the non-adiabatic pressure perturbation $\delta p_{\mathrm{nad}}$ defined in Eq.~(\ref{Eqn: delta p nad intro}). When the universe contains two barotropic perfect fluids, the term $\l[\delta \rho_{_1}/(\bar{\rho}_{_1} + \bar{p}_{_1})\,-\,\delta \rho_{_2}/(\bar{\rho}_{_2} + \bar{p}_{_2})\r]$ contributes to $\delta p_{\mathrm{nad}}$. However, when the matter components are canonical scalar fields, the term $\l[\delta q_{_1}/(\bar{\rho}_{_1} + \bar{p}_{_1})\,-\,\delta q_{_2}/(\bar{\rho}_{_2} + \bar{p}_{_2})\r]$ contributes to the non-adiabatic pressure perturbations in addition to the intrinsic entropy perturbations of the fields~\cite{Malik:2002jb}. Such isocurvature perturbations also arise in multi-field inflation~\cite{Wands:2007bd}.

Therefore, in a general two-field/fluid system, both isocurvature terms contribute to the total pressure perturbations. We thus expect the following relation for the total pressure perturbation:
\ber
\delta p = c_{e}^{2}\delta \rho +  3H\l(c_{a}^{2} - c_{e}^2\r) \delta q + \kappa_{_\rho}\l[\frac{\delta \rho_{_1}}{\bar{\rho}_{_1} + \bar{p}_{_1}} - \frac{\delta \rho_{_2}}{\bar{\rho}_{_2} + \bar{p}_{_2}}\r] +  \kappa_{_q}\l[\frac{\delta q_{_1}}{\bar{\rho}_{_1} + \bar{p}_{_1}} - \frac{\delta q_{_2}}{\bar{\rho}_{_2} + \bar{p}_{_2}}\r] \label{Eqn: relation delta rho p and q Two filed/fluid}
\eer
where $\kappa_{_\rho}$ and $\kappa_{_q}$ are background-dependent quantities, and $c_{a}^2 = \dot{\bar{p}}/\dot{\bar{\rho}}$ is the square of the adiabatic speed of sound for the total system, given by:
\beq
c_{a}^{2} = \frac{(\bar{\rho}_{_1} + \bar{p}_{_1})c_{a1}^{2} \,+\, (\bar{\rho}_{_2} + \bar{p}_{_2})c_{a2}^{2}}{\bar{\rho}+ \bar{p}}.\label{Eqn: ca2 adiabatic speed of sound two field/fluid}
\eeq
Note that Eq.~(\ref{Eqn: ca2 adiabatic speed of sound two field/fluid}) follows from Eqs.~(\ref{Eqn: background conservation eqn matter content 1}) and (\ref{Eqn: background conservation eqn matter content 2}), which assume that the two matter components are non-interacting.

If the two matter components are barotropic perfect fluids (with $\delta p_{_i} = c_{ai}^{2}\delta \rho_{_i}$), we expect the effective speed of sound $c_{e}^{2}$ to coincide with $c_{a}^{2}$. However, this is generally not the case when scalar fields are involved.

\subsection{The effective speed of sound in the case of a two fluid/field system}
In Eq.~(\ref{Eqn: relation delta rho p and q Two filed/fluid}), $c_{e}^{2}$ represents the square of the effective speed of sound of the total system. Deriving a closed-form expression for $c_{e}^{2}$ is a primary objective of this paper. It is straightforward to verify that Eq.~(\ref{Eqn: relation delta rho p and q Two filed/fluid}) is valid in every gauge of cosmological perturbations; that is, it is consistent with the gauge transformations provided in Eqs.~(\ref{Eqn: gauge transformation delta rho}) through (\ref{Eqn: gauge transformation delta q}). Furthermore, Eq.~(\ref{Eqn: relation delta rho p and q Two filed/fluid}) must be consistent with Eqs.~(\ref{Eqn: delta rho two field/fluid}) through (\ref{Eqn: TwoF relation delta rho p2 and q2}). To ensure this consistency, we substitute Eqs.~(\ref{Eqn: delta rho two field/fluid}) through (\ref{Eqn: TwoF relation delta rho p2 and q2}) into Eq.~(\ref{Eqn: relation delta rho p and q Two filed/fluid}) and equate the coefficients of $\delta \rho_{_1}$, $\delta \rho_{_2}$, $\delta q_{_1}$, and $\delta q_{_2}$ on both sides of the equation. Consequently, we find that:
\ber
c_{e}^{2} &=& \frac{(\bar{\rho}_{_1} + \bar{p}_{_1})c_{e1}^{2} \,+\, (\bar{\rho}_{_2} + \bar{p}_{_2})c_{e2}^{2}}{\bar{\rho}+ \bar{p}}\label{Eqn: ce2 effective speed of sound two field/fluid}\\
\kappa_{_\rho}  &=& \l(\frac{(\bar{\rho}_{_1} + \bar{p}_{_1})(\bar{\rho}_{_2} + \bar{p}_{_2})}{\bar{\rho}+ \bar{p}}\r)\l[c_{e1}^{2}\,-\,c_{e2}^{2}\r]\label{Eqn: c-rho  two field/fluid}\\
\kappa_{_q}  &=& 3H\l(\frac{(\bar{\rho}_{_1} + \bar{p}_{_1})(\bar{\rho}_{_2} + \bar{p}_{_2})}{\bar{\rho}+ \bar{p}}\r)\l[\l(c_{a1}^{2}\,-\,c_{e1}^{2}\r) \,-\,\l(c_{a2}^{2}\,-\,c_{e2}^{2}\r)\r]\label{Eqn: c-q  two field/fluid}
\eer

Eq.~(\ref{Eqn: ce2 effective speed of sound two field/fluid}) provides the required expression for the effective speed of sound of the total system. This becomes evident when Eq.~(\ref{Eqn: relation delta rho p and q Two filed/fluid}) is substituted into Eq.~(\ref{Eqn: GI Perturbed Einstein Eqn-3}) and  by further using Eqs.~(\ref{Eqn: GI Perturbed Einstein Eqn-1}) and (\ref{Eqn: GI Perturbed Einstein Eqn-2}), the evolution equation for the gauge-invariant Bardeen potential $\Phi$ is found to be:
\ber
& & \ddot{\Phi} + H\l[4 + 3c_a^2\r]\dot{\Phi} + \l[2\dot{H} + 3H^{2}(1 + c_a^2)\r]\Phi 
+ \l(\f{c_e^2 k^2}{a^2}\r)\Phi\nonumber \\
& & ~~~~~~ = 4 \pi G \kappa_{_\rho}\l[\frac{\delta \rho_{_1}}{\bar{\rho}_{_1} + \bar{p}_{_1}}\,-\,\frac{\delta \rho_{_2}}{\bar{\rho}_{_2} + \bar{p}_{_2}}\r] 
\,+\, 4 \pi G \kappa_{_q}\l[\frac{\delta q_{_1}}{\bar{\rho}_{_1} + \bar{p}_{_1}}\,-\,\frac{\delta q_{_2}}{\bar{\rho}_{_2} + \bar{p}_{_2}}\r].\label{Eqn: TF Phi dot dot Eqn}
\eer
Consequently, on scales much smaller than the Hubble radius, \textit{i.e.}, in the large-$k$ limit where $c_e k \gg aH$, Eq.~(\ref{Eqn: TF Phi dot dot Eqn}) implies that $\ddot{\Phi}\, \simeq \, - \l(c_e^2 k^2/a^2\r)\Phi$. Therefore, in a universe whose dynamics are driven by a two fluid/field system, the expression for $c_e$ obtained in Eq.~(\ref{Eqn: ce2 effective speed of sound two field/fluid}) is indeed the speed of sound with which the Bardeen potential $\Phi$ propagates in the limit $c_e k \gg aH$.

Furthermore, since the total comoving density perturbation $\delta \rho^{^{(gi)}}_{_c}$, defined in Eq.~(\ref{Eqn: GI rho in comoing gauge}), satisfies the Poisson equation~(\ref{Eqn: Poisson eqn for Phi}), it follows from Eq.~(\ref{Eqn: TF Phi dot dot Eqn}) that:
\ber
& &\ddot{\delta \rho}^{^{(gi)}}_{_c}+H\l[8 + 3c_a^2 \r]\dot{\delta \rho}^{^{(gi)}}_{_c} + \l[4\dot{H}+ 3H^{2}(5 + 3c_a^2)\r]\delta \rho^{^{(gi)}}_{_c}
+ \l(\f{c_e^2 k^2}{a^2}\r)\delta \rho^{^{(gi)}}_{_c}\nonumber \\
& & ~~~~~~  = -\frac{\kappa_{_\rho}k^{2}}{a^{2}}\l[\frac{\delta \rho_{_1}}{\bar{\rho}_{_1} + \bar{p}_{_1}}\,-\,\frac{\delta \rho_{_2}}{\bar{\rho}_{_2} + \bar{p}_{_2}}\r]
-\frac{\kappa_{_q}k^{2}}{a^{2}}\l[\frac{\delta q_{_1}}{\bar{\rho}_{_1} + \bar{p}_{_1}}\,-\,\frac{\delta q_{_2}}{\bar{\rho}_{_2} + \bar{p}_{_2}}\r]. \label{Eqn: TF comiving delta rho dot dot Eqn}
\eer
In the large-$k$ limit, this equation implies that:
\ber
\ddot{\delta \rho}^{^{(gi)}}_{_c} \simeq  -\l(\f{c_e^2 k^2}{a^2}\r)\delta \rho^{^{(gi)}}_{_c}-\frac{\kappa_{_\rho}k^{2}}{a^{2}}\l[\frac{\delta \rho_{_1}}{\bar{\rho}_{_1} + \bar{p}_{_1}}- \frac{\delta \rho_{_2}}{\bar{\rho}_{_2} + \bar{p}_{_2}}\r]
 -\frac{\kappa_{_q}k^{2}}{a^{2}}\l[\frac{\delta q_{_1}}{\bar{\rho}_{_1} + \bar{p}_{_1}} - \frac{\delta q_{_2}}{\bar{\rho}_{_2} + \bar{p}_{_2}}\r].
\label{Eqn: TF comiving delta rho dot dot Eqn large k}
\eer
However, since $\delta \rho^{^{(gi)}}_{_c} \propto -k^{2}\Phi$, in the large-$k$ limit we can approximate:
\ber
c_e^2 \delta \rho^{^{(gi)}}_{_c}  +  \kappa_{_\rho}\l[\frac{\delta \rho_{_1}}{\bar{\rho}_{_1} + \bar{p}_{_1}}\,-\,\frac{\delta \rho_{_2}}{\bar{\rho}_{_2} + \bar{p}_{_2}}\r]
+\ \kappa_{_q}\l[\frac{\delta q_{_1}}{\bar{\rho}_{_1} + \bar{p}_{_1}}\,-\,\frac{\delta q_{_2}}{\bar{\rho}_{_2} + \bar{p}_{_2}}\r] \approx c_e^2 \delta \rho^{^{(gi)}}_{_c}.\nonumber
\eer
Consequently, it follows from Eq.~(\ref{Eqn: TF comiving delta rho dot dot Eqn large k}) that, in the large-$k$ limit, $\delta \rho^{^{(gi)}}_{_c}$ satisfies the wave equation $\ddot{\delta \rho}^{^{(gi)}}_{_c} \simeq -(k^{2}c_e^2/a^{2})\delta \rho^{^{(gi)}}_{_c}$, where $c_e^2$ is the speed of sound defined in Eq.~(\ref{Eqn: ce2 effective speed of sound two field/fluid}).

Note that in the case where the two matter components are barotropic perfect fluids, for which, by definition, $c_{e1}^{2} = c_{a1}^{2}$ and $c_{e2}^{2} = c_{a2}^{2}$, this effective speed of sound $c_{e}^{2}$ of the total system, defined in Eq.~(\ref{Eqn: ce2 effective speed of sound two field/fluid}), exactly coincides with the adiabatic speed of sound of the total system $c_{a}^{2}$ given in Eq.~(\ref{Eqn: ca2 adiabatic speed of sound two field/fluid}). This is also the case when the two matter components are pure kinetic scalar fields, which similarly yield $c_{e1}^{2} = c_{a1}^{2}$ and $c_{e2}^{2} = c_{a2}^{2}$, and consequently $c_{e}^{2} = c_{a}^{2}$ for the total system.

In the case where the two matter components are minimally coupled canonical scalar fields with Lagrangians $\mathcal{L}_{1}$ and $\mathcal{L}_{2}$, defined as:
\berstr
\mathcal{L}_{1}(\phi, \partial_\mu \phi) = \l(\frac{1}{2}\r)\partial_\mu \phi \partial^\mu \phi - V(\phi),\\
\mathrm{and}~~~~~~~~~~~~~~~~~~~~~~~~~~~~~~~~~~~~~~~~~~&&\\
\mathcal{L}_{2}(\varphi, \partial_\mu \varphi) = \l(\frac{1}{2}\r)\partial_\mu \varphi \partial^\mu \varphi - U(\varphi),
\eerstr
respectively, the effective speed of sound for each of the two canonical scalar fields is unity~\cite{Hu:1998kj}, \textit{i.e.}, $c_{e1}^{2} = c_{e2}^{2} = 1$. Consequently, from Eq.~(\ref{Eqn: ce2 effective speed of sound two field/fluid}), the effective speed of sound $c_{e}$ of the total system is also unity.

Now, if both scalar fields are non-canonical with Lagrangians of the type~\cite{Mukhanov:2005bu,Unnikrishnan:2008ki,Unnikrishnan:2012zu}:
\ber
\mathcal{L}_{1}(\phi, X) &=& X^\alpha - V(\phi)\label{Eqn: Lagrangian NC scalar field-1}\\
\mathcal{L}_{2}(\varphi, Y) &=& Y^\alpha - U(\varphi)\label{Eqn: Lagrangian NC scalar field-2}
\eer
where $\alpha$ is a constant, $X = (1/2)\partial_\mu \phi \partial^\mu \phi$, and $Y = (1/2)\partial_\mu \varphi \partial^\mu \varphi$, the effective speed of sound for each of these fields is a constant given by $c_{e1}^{2} = c_{e2}^{2} = (2\alpha - 1)^{-1}$. Consequently, from Eq.~(\ref{Eqn: ce2 effective speed of sound two field/fluid}), the square of the effective speed of sound for the total system is also $c_e^2 = (2\alpha - 1)^{-1}$.

In general, if we consider two minimally coupled non-canonical scalar fields with Lagrangians $\mathcal{F}(\phi, X)$ and $\mathcal{G}(\varphi, Y)$, where $X = (1/2)\partial_\mu \phi \partial^\mu \phi$ and $Y = (1/2)\partial_\mu \varphi \partial^\mu \varphi$, it follows from Eq.~(\ref{Eqn: ce2 effective speed of sound two field/fluid}) that the effective speed of sound of the total system is:
\beq
c_{e}^{2} = \frac{X\mathcal{F}_{X}^2(\mathcal{G}_Y + 2Y\mathcal{G}_{YY}) + Y\mathcal{G}_{Y}^2(\mathcal{F}_X + 2X\mathcal{F}_{XX})}{(X\mathcal{F}_X + Y\mathcal{G}_Y)(\mathcal{F}_X + 2X\mathcal{F}_{XX})(\mathcal{G}_Y + 2Y\mathcal{G}_{YY})}.\label{Eqn: c_e^2 for two NC scalar fields}
\eeq
This expression generalizes the definition of the effective speed of sound for a single non-canonical scalar field provided in Eq.~(\ref{Eqn: c_e^2 for NC scalar field}), originally introduced by Garriga and Mukhanov in Ref.~\cite{Garriga:1999vw}, to the case of two minimally coupled non-canonical scalars. Eq.~(\ref{Eqn: c_e^2 for two NC scalar fields}) constitutes one of the main results of this paper.
Note that the above expression for $c_{e}^{2}$ is valid only for a system consisting of two non-interacting non-canonical scalar fields with separable Lagrangians.

Note that Eq.~(\ref{Eqn: ce2 effective speed of sound two field/fluid}) can be expressed as:
\ber
c_{e}^{2} = c_{e1}^{2} - \l[\frac{(c_{e1}^{2} -c_{e2}^{2})(\bar{\rho}_{_2} + \bar{p}_{_2})}{\bar{\rho} + \bar{p}}\r]
 = c_{e2}^{2} + \l[\frac{(c_{e1}^{2} -c_{e2}^{2})(\bar{\rho}_{_1} + \bar{p}_{_1})}{\bar{\rho} + \bar{p}}\r].
\label{Eqn: c_e^2 TF evolution eqn}
\eer
It is evident from the above equation that if $c_{e1}^{2} > c_{e2}^{2}$, then $c_{e2}^{2} \leq c_{e}^{2} \leq c_{e1}^{2}$. Consequently, $0 \leq c_{e}^{2} \leq 1$, provided that for both matter components $0 \leq c_{e1}^{2} \leq 1$ and $0 \leq c_{e2}^{2} \leq 1$, as required by the causality condition which forbids speeds of sound greater than the speed of light. In this paper, we adopt units where the speed of light $c = 1$. Therefore, if $c_{e1}$ and $c_{e2}$ are subluminal, \textit{i.e.}, $c_{e1}^{2} \leq 1$ and $c_{e2}^{2} \leq 1$, it follows from Eq.~(\ref{Eqn: c_e^2 TF evolution eqn}) that the total effective speed of sound $c_{e}$ is also subluminal. In a scenario where the dynamics of the universe are such that the component $\bar{\rho}_{_1}$ dominates initially, followed by an epoch where $\bar{\rho}_{_2}$ dominates, Eq.~(\ref{Eqn: c_e^2 TF evolution eqn}) implies that $c_{e}^{2}$ evolves from an initial value of $c_{e1}^{2}$ and asymptotically approaches $c_{e2}^{2}$.

\subsection{Non-adiabatic pressure perturbation in the two field/fluid case}
Using Eqs.~(\ref{Eqn: GI Perturbed Einstein Eqn-1}), (\ref{Eqn: GI Perturbed Einstein Eqn-2}), and (\ref{Eqn: relation delta rho p and q Two filed/fluid}), the non-adiabatic pressure perturbation, defined as $\delta p_{\mathrm{nad}} = \delta p - c_{a}^2\,\delta \rho$, is found to be:
\ber
\delta p_{\mathrm{nad}} = \l(\frac{c_{a}^2 - c_{e}^2}{4 \pi G a^2}\r)k^2\Phi + \kappa_{_\rho}\l[\frac{\delta \rho_{_1}}{\bar{\rho}_{_1} + \bar{p}_{_1}}\,-\,\frac{\delta \rho_{_2}}{\bar{\rho}_{_2} + \bar{p}_{_2}}\r] + \kappa_{_q}\l[\frac{\delta q_{_1}}{\bar{\rho}_{_1} + \bar{p}_{_1}}\,-\,\frac{\delta q_{_2}}{\bar{\rho}_{_2} + \bar{p}_{_2}}\r].\label{Eqn: delta p nad Two field/fluid}
\eer
Consequently, the two isocurvature perturbation terms, namely $[\delta \rho_{_1}/(\bar{\rho}_{_1} + \bar{p}_{_1}) - \delta \rho_{_2}/(\bar{\rho}_{_2} + \bar{p}_{_2})]$ and $[\delta q_{_1}/(\bar{\rho}_{_1} + \bar{p}_{_1}) - \delta q_{_2}/(\bar{\rho}_{_2} + \bar{p}_{_2})]$, contribute to $\delta p_{\mathrm{nad}}$. Therefore, on super-horizon scales, \textit{i.e.}, in the limit $k \rightarrow 0$, $\delta p_{\mathrm{nad}}$ may not vanish due to these isocurvature terms. This stands in contrast to the single-fluid or single-field case, where $\delta p_{\mathrm{nad}} \rightarrow 0$ as $k \rightarrow 0$, as evidenced by Eq.~(\ref{Eqn: delta p nad eqn single field k dependence}).

Following Ref.~\cite{Malik:2004tf}, the total $\delta p_{\mathrm{nad}}$ given in Eq.~(\ref{Eqn: delta p nad Two field/fluid}) can be decomposed as:
\ber
\delta p_{\mathrm{nad}} = \delta p_{\mathrm{intr}} + \delta p_{\mathrm{rel}},\label{Eqn: delta p nad decomposition}
\eer 
where $\delta p_{\mathrm{intr}}$ is the sum of the intrinsic non-adiabatic pressure perturbation of each matter component, i.e.,
\ber
\delta p_{\mathrm{intr}} = \delta p_{\mathrm{nad1}} + \delta p_{\mathrm{nad2}} =    \l(\delta p_{_1}  - c_{a1}^2\,\delta \rho_{_1}\r)     +  \l(\delta p_{_2}  - c_{a2}^2\,\delta \rho_{_2}\r),\label{Eqn: delta p intr as sum}
\eer
and $\delta p_{\mathrm{rel}}$ is the relative non-adiabatic pressure perturbation. 
Using Eqs.~(\ref{Eqn: TwoF relation delta rho p1 and q1}) and (\ref{Eqn: TwoF relation delta rho p2 and q2}) along with the perturbed Einstein's equations (\ref{Eqn: GI Perturbed Einstein Eqn-1}) and (\ref{Eqn: GI Perturbed Einstein Eqn-2}), it turns out that 
\ber
\delta p_{\mathrm{intr}} = \l(\frac{c_{a}^2 - c_{e}^2}{4 \pi G a^2}\r)k^2\Phi 
 - \frac{\kappa_{_q}}{3H}\l[\l(\frac{\delta \rho_{_1}}{\bar{\rho}_{_1} + \bar{p}_{_1}}\,-\,\frac{\delta \rho_{_2}}{\bar{\rho}_{_2} + \bar{p}_{_2}}\r)
- 3H \l(\frac{\delta q_{_1}}{\bar{\rho}_{_1} + \bar{p}_{_1}}\,-\,\frac{\delta q_{_2}}{\bar{\rho}_{_2} + \bar{p}_{_2}}\r)
\r],\label{Eqn: delta p intr}
\eer
where $\kappa_{_q}$ is defined in Eq.~(\ref{Eqn: c-q  two field/fluid}).
Consequently, using Eqs.~(\ref{Eqn: c-rho  two field/fluid}) and (\ref{Eqn: c-q  two field/fluid}), it follows from Eqs.~(\ref{Eqn: delta p nad Two field/fluid}) and  (\ref{Eqn: delta p nad decomposition}) that~\cite{Malik:2004tf}
\ber
\delta p_{\mathrm{rel}} = \l(c_{a1}^2 - c_{a2}^2\r)\l[\frac{(\bar{\rho}_{_1} + \bar{p}_{_1})(\bar{\rho}_{_2} + \bar{p}_{_2})}{\bar{\rho}+ \bar{p}}\r]\l(\frac{\delta \rho_{_1}}{\bar{\rho}_{_1} + \bar{p}_{_1}}\,-\,\frac{\delta \rho_{_2}}{\bar{\rho}_{_2} + \bar{p}_{_2}}\r)
\label{Eqn: delta p rel}
\eer

Note that when the two matter components are barotropic perfect fluids or two pure-kinetic non-canonical scalar fields, for which $c_{e1}^{2} = c_{a1}^{2}$ and $c_{e2}^{2} = c_{a2}^{2}$, it is evident from Eq.~(\ref{Eqn: c-q  two field/fluid}) that $\kappa_{_q} = 0$. In this scenario, only the isocurvature perturbation term $[\delta \rho_{_1}/(\bar{\rho}_{_1} + \bar{p}_{_1})\,-\,\delta \rho_{_2}/(\bar{\rho}_{_2} + \bar{p}_{_2})]$ contributes to $\delta p_{\mathrm{nad}}$ described in Eq.~(\ref{Eqn: delta p nad Two field/fluid}).
Since $c_{e1}^{2} = c_{a1}^{2}$ and $c_{e2}^{2} = c_{a2}^{2}$ implies that $c_{e}^2  = c_{a}^2$ from Eqs.~(\ref{Eqn: ca2 adiabatic speed of sound two field/fluid}) and (\ref{Eqn: ce2 effective speed of sound two field/fluid}), and $\kappa_{_q} = 0$ from Eq.~(\ref{Eqn: c-q  two field/fluid}), it follows from Eq.~(\ref{Eqn: delta p intr}) that $\delta p_{\mathrm{intr}} = 0$. 
Consequently, in such a system of two barotropic perfect fluids or two pure-kinetic non-canonical scalar fields, the total non-adiabatic pressure perturbation $\delta p_{\mathrm{nad}} = \delta p_{\mathrm{rel}}$ given in Eq.~(\ref{Eqn: delta p rel}).

However, if the dynamics of the universe are driven by two non-canonical scalars with Lagrangians of the type described in Eqs.~(\ref{Eqn: Lagrangian NC scalar field-1}) and (\ref{Eqn: Lagrangian NC scalar field-2}), for which $c_{e1}^{2} = c_{e2}^{2}$, it follows from Eq.~(\ref{Eqn: c-rho  two field/fluid}) that $\kappa_{_\rho} = 0$. Therefore, the density-related isocurvature perturbation term $[\delta \rho_{_1}/(\bar{\rho}_{_1} + \bar{p}_{_1})\,-\,\delta \rho_{_2}/(\bar{\rho}_{_2} + \bar{p}_{_2})]$ does not contribute to $\delta p_{\mathrm{nad}}$. Consequently, in the two-field case with Lagrangians given by Eqs.~(\ref{Eqn: Lagrangian NC scalar field-1}) and (\ref{Eqn: Lagrangian NC scalar field-2}), the non-adiabatic pressure perturbation can be expressed as:
\beq
\delta p_{\mathrm{nad}} \,=\,\l[\frac{c_{a}^2 - (2\alpha-1)^{-1}}{4 \pi G a^2}\r]k^2\Phi \;+\; \kappa_{_q}\l[\frac{\delta \psi}{\dot{\bar{\psi}}} \,-\, \frac{\delta \phi}{\dot{\bar{\phi}}}\r].
\label{Eq: delat p nad two NC with constant speed}
\eeq
To arrive at the above equation, we have utilized the definition of $\delta q$ from Eq.~(\ref{Eqn: Definition of delta q}) and its non-canonical form in Eq.~(\ref{Eqn: perturbed delta q NC scalar field}). 
Note that in this case both $\delta p_{\mathrm{intr}}$ and $\delta p_{\mathrm{rel}}$ contribute to the total non-adiabatic pressure perturbation $\delta p_{\mathrm{nad}}$.

In the most general case, where the dynamics are driven by two minimally coupled non-canonical scalar fields that are not pure-kinetic and for which $c_{e1}^{2} \neq c_{e2}^{2}$, all three terms on the right-hand side of Eq.~(\ref{Eqn: delta p nad Two field/fluid}) contribute to $\delta p_{\mathrm{nad}}$. 
As in the previous case, both $\delta p_{\mathrm{intr}}$ and $\delta p_{\mathrm{rel}}$
contribute to $\delta p_{\mathrm{nad}}$. 
Note that in this paper we have considered a system consisting of non-interacting matter components and/or non-canonical scalar fields with separable Lagrangians wherein there is no direct energy-momentum transfer between the matter components. Ref.~\cite{Malik:2004tf} considered the case of energy-momentum transfer between the fluids and scalar fields with non-separable potentials.

\section{\label{Sec: multi fluid/field}The case of a multi-field/fluid cosmological system}
Let us now consider a scenario where the dynamics of the universe are driven by a collection of minimally coupled matter components like barotropic perfect fluids and scalar fields (which may be either canonical or non-canonical). In this section, we extend the definition of the effective speed of sound for the total system to such a multi-component universe.

Let us assume that there are $n$ minimally coupled matter components in the universe, which may include barotropic perfect fluids, canonical scalar fields, and non-canonical scalar fields. In a background FLRW universe with the line element given by Eq.~(\ref{Eqn: FRW line element}), the total energy density $\bar{\rho}$ and pressure $\bar{p}$ are given by:
\ber
\bar{\rho} =  \sum_{i=1}^{n} \bar{\rho}_{_i}, \label{Eqn: rho total multi field/fluid}\\
\bar{p} =  \sum_{i=1}^{n} \bar{p}_{_i}, \label{Eqn: p total multi field/fluid}
\eer
where $\bar{\rho}_{_i}$ and $\bar{p}_{_i}$ denote the energy density and pressure of the $i^{\mathrm{th}}$ component, respectively. The evolution of the scale factor $a(t)$ is governed by the following set of $n + 1$ equations:
\ber
\frac{\dot{a}^2}{a^2} &=& \l(\frac{8 \pi G }{3}\r)\sum_{i=1}^{n} \bar{\rho}_{_i},\label{Eqn: Friedmann eqn multifield}\\
\dot{\bar{\rho}}_{_i} &=& -3H(\bar{\rho}_{_i} + \bar{p}_{_i}).\label{Eqn: Conservation Eqn multifield}
\eer
Similarly, the variables describing the matter perturbations of the total system, \textit{viz.}, $\delta \rho$, $\delta p$, and $\delta q$, are obtained by summing the individual contributions:
\ber
\delta \rho \,&=&\,\sum_{i=1}^{n}\delta \rho_{_i},\label{Eqn: delta rho multi field/fluid}\\
\delta p\,&=&\,\sum_{i=1}^{n}\delta p_{_i},\label{Eqn: delta p multi field/fluid}\\
\delta q \,&=&\,\sum_{i=1}^{n}\delta q_{_i}.\label{Eqn: delta q multi field/fluid}
\eer

As argued in the preceding sections, at linear order in the perturbations, the pressure perturbation $\delta p_{_i}$ for each matter component is linearly related to its energy density perturbation $\delta \rho_{_i}$ and $\delta q_{_i}$ by:
\beq
\delta p_{_i} = c_{ei}^{2}\,\delta \rho_{_i} +  3H\l(c_{ai}^{2} - c_{ei}^2\r) \delta q_{_i},\label{Eqn: multiF relation delta rho pi and qi}
\eeq
where $c_{ei}$ is the effective speed of sound of the $i^{\mathrm{th}}$ matter component and $c_{ai} = \sqrt{\dot{\bar{p}}_{_i}/\dot{\bar{\rho}}_{_i}}$ is the adiabatic speed of sound of that same component. For barotropic perfect fluids, $c_{ei}^{2} = c_{ai}^{2}$, whereas for general non-canonical scalar fields with $\mathcal{L}(X,\phi)$, the effective speed of sound is given by Eq.~(\ref{Eqn: c_e^2 for NC scalar field}).

As discussed in Sec.~\ref{Sec: single fluid/field}, Eq.~(\ref{Eqn: multiF relation delta rho pi and qi}) follows from the requirement that the linear dependence of $\delta p_{_i}$ on $\delta \rho_{_i}$ and $\delta q_{_i}$ must be independent of the gauge choice. This implies that in every gauge where $\delta \rho_{_i} \neq 0$ and $\delta q_{_i} \neq 0$, the pressure perturbation $\delta p_{_i}$ must relate to $\delta \rho_{_i}$ and $\delta q_{_i}$ exactly as described in Eq.~(\ref{Eqn: multiF relation delta rho pi and qi}). We, therefore, expect a similar relation between the total variables $\delta p$, $\delta \rho$, and $\delta q$ in a multi-field/fluid universe. However, as noted in Sec.~\ref{Sec: two fluid/field}, gauge-invariant isocurvature (or entropy) perturbation terms, such as $[\delta\rho_{_i}/(\bar{\rho}_{_i} + \bar{p}_{_i}) - \delta\rho_{_j}/(\bar{\rho}_{_j} + \bar{p}_{_j})]$ and $[\delta q_{_i}/(\bar{\rho}_{_i} + \bar{p}_{_i}) - \delta q_{_j}/(\bar{\rho}_{_j} + \bar{p}_{_j})]$, also contribute to the total pressure perturbation $\delta p$. These terms specifically contribute to the non-adiabatic pressure perturbation $\delta p_{\mathrm{nad}}$. Consequently, we can express the total pressure perturbation $\delta p$ as:
\beq
\delta p \,= c_{e}^{2}\,\delta \rho +  3H\l(c_{a}^{2} - c_{e}^2\r) \delta q + \sum_{i,j=1}^{n}\l[a_{ij}\delta s_{ij} \,+\, b_{ij}\delta\sigma_{ij}\r], \label{Eqn: relation delta rho p and q multi filed/fluid}
\eeq
where 
\ber
\delta s_{ij} \,&=&\,(\bar{\rho} + \bar{p})\l[\frac{\delta \rho_{_i}}{\bar{\rho}_{_i} + \bar{p}_{_i}}\,-\,\frac{\delta \rho_{_j}}{\bar{\rho}_{_j} + \bar{p}_{_j}}\r],\label{Eqn: delta s ij}\\
\delta\sigma_{ij} \,&=&\, 3H(\bar{\rho} + \bar{p})\l[\frac{\delta q_{_i}}{\bar{\rho}_{_i} + \bar{p}_{_i}}\,-\,\frac{\delta q_{_j}}{\bar{\rho}_{_j} + \bar{p}_{_j}}\r].\label{Eqn: delta sigma ij}
\eer
The gauge-invariant isocurvature terms $\delta s_{ij}$ and $\delta\sigma_{ij}$ possess the same dimensions as $\delta p$ in Eq.~(\ref{Eqn: relation delta rho p and q multi filed/fluid}), rendering the coefficients $a_{ij}$ and $b_{ij}$ dimensionless, background-dependent quantities. Since $\delta s_{ij} = - \delta s_{ji}$ and $\delta\sigma_{ij} = -\delta\sigma_{ji}$, the coefficients $a_{ij}$ and $b_{ij}$ in Eq.~(\ref{Eqn: relation delta rho p and q multi filed/fluid}) are taken to be anti-symmetric. For $n = 2$, Eq.~(\ref{Eqn: relation delta rho p and q multi filed/fluid}) reduces to the two-field case in Eq.~(\ref{Eqn: relation delta rho p and q Two filed/fluid}) with $\kappa_{_\rho} = 2a_{12}(\bar{\rho} + \bar{p})$ and $\kappa_{_q} = 2b_{12}(\bar{\rho} + \bar{p})$.

In Eq.~(\ref{Eqn: relation delta rho p and q multi filed/fluid}), $c_{a}^{2} = \dot{\bar{p}}/\dot{\bar{\rho}}$ represents the square of the adiabatic speed of sound for the total system. Therefore, using Eqs.~(\ref{Eqn: rho total multi field/fluid}) and (\ref{Eqn: p total multi field/fluid}), it follows that~\cite{Malik:2004tf}:
\beq
c_{a}^{2} =  \l(\frac{1}{\bar{\rho} + \bar{p}}\r)\sum_{i=1}^{n}\l[(\bar{\rho}_{_i} + \bar{p}_{_i})c_{ai}^2\r]. \label{Eqn: ca2 total in multi field/fluid}
\eeq

In Eq.~(\ref{Eqn: relation delta rho p and q multi filed/fluid}), $c_{e}$ represents the effective speed of sound of the total multi-field/fluid system. Deriving a closed-form expression for $c_{e}$ is the primary objective of this section. Note that under an infinitesimal coordinate transformation $x^{\mu} \rightarrow \tilde{x}^{\mu} = x^{\mu} + \xi^{\mu}$, the perturbation variables transform as $\delta \rho \rightarrow \widetilde{\delta \rho}$, $\delta p \rightarrow \widetilde{\delta p}$, and $\delta q \rightarrow \widetilde{\delta q}$. It is straightforward to verify that these transformed quantities relate to one another in precisely the same manner as shown in Eq.~(\ref{Eqn: relation delta rho p and q multi filed/fluid}). Consequently, the $c_{e}^2$ defined in Eq.~(\ref{Eqn: relation delta rho p and q multi filed/fluid}) is a gauge-invariant quantity.

Substituting Eqs.~(\ref{Eqn: delta rho multi field/fluid}) through (\ref{Eqn: delta q multi field/fluid}) into Eq.~(\ref{Eqn: relation delta rho p and q multi filed/fluid}), we obtain:
\ber
\sum_{i=1}^{n}\delta p_{_i} \,= c_{e}^{2}\,\sum_{i=1}^{n}\delta \rho_{_i} +  3H\l(c_{a}^{2} - c_{e}^2\r) \sum_{i=1}^{n}\delta q_{_i}
+ \sum_{i,j = 1}^{n}\l[a_{ij}\delta s_{ij} \,+\, b_{ij}\delta\sigma_{ij}\r]. \label{Eqn: relation-2 delta rho p and q multi field/fluid}
\eer
For consistency, this relation must hold for any configuration of the individual components as described by Eq.~(\ref{Eqn: multiF relation delta rho pi and qi}). Therefore, by substituting Eq.~(\ref{Eqn: multiF relation delta rho pi and qi}) into the left-hand side and equating the coefficients of $\delta \rho_{_i}$ and $\delta q_{_i}$ on both sides, we find:
\ber
c_{e}^{2} &=&  \l(\frac{1}{\bar{\rho} + \bar{p}}\r)\sum_{i=1}^{n}\l[(\bar{\rho}_{_i} + \bar{p}_{_i})c_{ei}^2\r], \label{Eqn: ce2 total in multi field/fluid}\\
a_{ij}  &=& \l[\frac{(\bar{\rho}_{_i} + \bar{p}_{_i})(\bar{\rho}_{_j} + \bar{p}_{_j})}{2(\bar{\rho}+ \bar{p})^2}\r]\l(c_{ei}^{2}\,-\,c_{ej}^{2}\r), \label{Eqn: aij multi field/fluid}\\
b_{ij}  &=& \l[\frac{(\bar{\rho}_{_i} + \bar{p}_{_i})(\bar{\rho}_{_j} + \bar{p}_{_j})}{2(\bar{\rho}+ \bar{p})^2}\r]\l[\l(c_{ai}^{2}\,-\,c_{ei}^{2}\r) \,-\,\l(c_{aj}^{2}\,-\,c_{ej}^{2}\r)\r]. \label{Eqn: bij multi field/fluid}
\eer

In a multi-field/fluid system with $\delta p$ given by Eq.~(\ref{Eqn: relation delta rho p and q multi filed/fluid}), the evolution equation for the gauge-invariant Bardeen potential $\Phi$ is found to be:
\ber
\ddot{\Phi} + H\l[4 + 3c_a^2\r]\dot{\Phi} + \l[2\dot{H} + 3H^{2}(1 + c_a^2)\r]\Phi 
+ \l(\f{c_e^2 k^2}{a^2}\r)\Phi 
= 4 \pi G \sum_{i,j = 1}^{n}\l[a_{ij}\delta s_{ij} + b_{ij}\delta\sigma_{ij}\r]. \label{Eqn: multi field Phi dot dot Eqn}
\eer
Consequently, in the large-$k$ limit, this reduces to $\ddot{\Phi} \simeq - (c_e^2 k^2/a^2)\Phi$. Similarly, the gauge-invariant comoving density perturbation $\delta \rho^{^{(gi)}}_{_c}$ of the total system, defined in Eq.~(\ref{Eqn: GI rho in comoing gauge}), satisfies the same wave equation $\ddot{\delta \rho}^{^{(gi)}}_{_c} \simeq -(k^{2}c_e^2/a^{2})\delta \rho^{^{(gi)}}_{_c}$ in the large-$k$ limit. It is therefore evident that $c_{e}$, as defined in Eq.~(\ref{Eqn: ce2 total in multi field/fluid}), is indeed the effective speed of sound of the total system when the dynamics of the universe are driven by a collection of minimally coupled canonical scalar fields, non-canonical scalar fields, and barotropic perfect fluids. 
Eq.~(\ref{Eqn: ce2 total in multi field/fluid}) constitutes the central result of this paper. The effective speed of sound defined here is not only gauge-invariant but also independent of scale $k$. Being a purely background-dependent quantity, it can be treated as a robust parameter to quantify and classify perturbations within such multi-component cosmological systems.

Note that in Eq.~(\ref{Eqn: ce2 total in multi field/fluid}), $c_{ei}^2$ is the square of the effective speed of sound for the $i^{\mathrm{th}}$ matter component. For barotropic perfect fluids, $c_{ei}^2 = c_{ai}^2 = \dot{\bar{p}}_{_i}/\dot{\bar{\rho}}_{_i}$. Similarly, for pure-kinetic non-canonical scalar field models with Lagrangian $\mathcal{L}(X)$, it is found that $c_{ei}^2 = c_{ai}^2$. Comparing Eq.~(\ref{Eqn: ca2 total in multi field/fluid}) with Eq.~(\ref{Eqn: ce2 total in multi field/fluid}) reveals that, when the dynamics of the universe are driven by a collection of minimally coupled pure-kinetic non-canonical scalar fields, the effective speed of sound $c_{e}$ of the total system is identically equal to the adiabatic speed of sound $c_{a}$. 
Consequently, the evolution of the scale factor $a(t)$, as determined by Eqs.~(\ref{Eqn: Friedmann eqn multifield}) and (\ref{Eqn: Conservation Eqn multifield}), and the gauge-invariant Bardeen potential $\Phi$, as determined by Eq.~(\ref{Eqn: multi field Phi dot dot Eqn}), are identical to those in a system comprising an equivalent collection of barotropic perfect fluids. This result generalizes the known equivalence between a single pure-kinetic non-canonical scalar field and a barotropic perfect fluid~\cite{Unnikrishnan:2010ag,Arroja:2010wy} to the case where multiple pure-kinetic non-canonical scalar fields are equivalent to a collection of barotropic perfect fluids.

In this paper, we have defined $c_{e}$, as given in Eq.~(\ref{Eqn: ce2 total in multi field/fluid}), as the effective speed of sound for the total system when the universe contains a collection of minimally coupled canonical scalar fields, non-canonical scalar fields, and barotropic perfect fluids. This definition is justified by the fact that, on small scales, the gauge-invariant Bardeen potential $\Phi$ satisfies a wave equation with $c_{e}$ acting as the speed of propagation, as evidenced by Eq.~(\ref{Eqn: multi field Phi dot dot Eqn}). 
However, it is important to note that in Ref.~\cite{Romano:2018frb}, the effective speed of sound in multi-field systems is defined differently. In that work, it is defined as the ratio of $\delta p$ to $\delta \rho$ specifically in the comoving gauge (the gauge where $\delta q = 0$, with $\delta q$ defined as in Eq.~(\ref{Eqn: Defnition delta T 0i})). Denoting this ratio as $v_{s}^2(t, x^{i})$ following Ref.~\cite{Romano:2018frb}, it follows from Eq.~(\ref{Eqn: relation delta rho p and q multi filed/fluid}) that, in the comoving gauge:
\beq
\l(v_{s}^2 - c_{e}^2\r)\delta \rho = \sum_{i,j=1}^{n}\l[a_{ij}\delta s_{ij} \,+\, b_{ij}\delta\sigma_{ij}\r].
\label{Eqn: ce and vs reln}
\eeq
The above equation implies that $v_{s}^2 \approx c_{e}^2$ only when the contributions from the isocurvature perturbation terms $\delta s_{ij}$ and $\delta\sigma_{ij}$ are negligible. Otherwise, unlike our definition of $c_{e}^2$, the quantity $v_{s}^2$ remains scale-dependent and gauge-specific.

Note that the speed of sound for cosmological perturbations is physically relevant as the propagation speed of the metric perturbation, specifically the Bardeen potential $\Phi$, primarily at small scales or in the large-$k$ limit. As it is generally expected that isocurvature perturbation terms such as $\delta s_{ij}$ and $\delta\sigma_{ij}$ do not significantly influence the evolution of perturbations at these small scales, the values for the speed of propagation obtained from the two definitions of the effective speed of sound, \emph{viz}.\ Eq.~(\ref{Eqn: ce2 total in multi field/fluid}) and the one defined in the one defined in Ref.~\cite{Romano:2018frb}, coincide in this small scale limit. 
However, the primary advantage of utilizing Eq.~(\ref{Eqn: ce2 total in multi field/fluid}) is that $c_e$ is a scale-independent quantity, expressed purely in terms of background variables such as $\bar{\rho}_i$ and $\bar{p}_i$. Nevertheless, the space-dependent speed of sound defined in Ref.~\cite{Romano:2018frb} maybe advantageous for expressing perturbations in more general systems involving non-minimally coupled matter components or within the framework of modified gravity. The closed-form expression for $c_{e}$ derived in Eq.~(\ref{Eqn: ce2 total in multi field/fluid}) provides the total effective speed of sound for cosmological perturbations when the constituent matter components, such as canonical and non-canonical scalar fields and barotropic perfect fluids, are minimally coupled to one another.

\section{Curvature Perturbation}\label{Sec: Curvature perturbation}
Using Eqs.~(\ref{Eqn: gauge transformation psi}) and (\ref{Eqn: gauge transformation delta q}), we can construct the following gauge-invariant quantity~\cite{Baumann:2009ds}:
\beq
\mathcal{R} = \psi -\frac{H\delta q}{\bar{\rho} + \bar{p}}.\label{Eqn: Curvature Perturbation}
\eeq
This quantity $\mathcal{R}$ is known as the comoving curvature perturbation since, in the comoving gauge (where $\delta q = 0$), $\mathcal{R} = \psi$. Here, $\psi$ is proportional to the curvature of the 3-space of the perturbed FLRW line element given by Eq.~(\ref{Eqn: perturbed FRW line element})~\cite{Riotto:2002yw}. Using Eqs.~(\ref{Eqn: Bardeen Potential Psi}) and (\ref{Eqn: gauge invarian delta q}), we can also express $\mathcal{R}$ as a linear combination of gauge-invariant quantities:
\beq
\mathcal{R} = \Phi -\frac{\,H\delta q^{^{(gi)}}_{_B}}{\bar{\rho} + \bar{p}},\label{Eqn: Curvature Perturbation in GI terms}
\eeq
where $\Phi$ is the gauge-invariant Bardeen potential and $\delta q^{^{(gi)}}_{_B}$ is the corresponding gauge-invariant momentum density perturbation. Note that in the above expression, we assume that no matter component contributes to the anisotropic stress in the energy-momentum tensor; consequently, we have utilized the identity in Eq.~(\ref{Eqn: equality of Bardeen Potentials}).

Taking the time derivative of Eq.~(\ref{Eqn: Curvature Perturbation in GI terms}) and using Eqs.~(\ref{Eqn: GI Perturbed Einstein Eqn-2}) and (\ref{Eqn: GI delta q conservation eqn}), we obtain~\cite{Mukhanov:1990me,Malik:2008im}:
\beq
\dot{\mathcal{R}} = -\l(\frac{H\,c_{a}^{2}}{\bar{\rho} + \bar{p}}\r)\l(\frac{k^2\Phi}{4 \pi G a^{2}}\r) \,+\, \l(\frac{H}{\bar{\rho} + \bar{p}}\r)\delta p_{\mathrm{nad}}.\label{Eqn: Curvature Perturbation R dot-1}
\eeq
This equation is valid for both single-field/fluid and multi-field/fluid models. Note that $c_{a}^{2}$ in Eq.~(\ref{Eqn: Curvature Perturbation R dot-1}) denotes the square of the adiabatic speed of sound for the total system. For a single barotropic perfect fluid, the non-adiabatic pressure perturbation $\delta p_{\mathrm{nad}}$ vanishes; therefore, $\dot{\mathcal{R}} \rightarrow 0$ as $k \rightarrow 0$, implying that $\mathcal{R}$ is conserved on super-horizon scales~\cite{Mukhanov:1990me,Malik:2008im}.

In the general case, using Eqs.~(\ref{Eqn: GI Perturbed Einstein Eqn-1}), (\ref{Eqn: GI Perturbed Einstein Eqn-2}), and (\ref{Eqn: relation delta rho p and q multi filed/fluid}), we can express the non-adiabatic pressure perturbation, defined as $\delta p_{\mathrm{nad}} = \delta p - c_{a}^2\,\delta \rho$, as:
\beq
\delta p_{\mathrm{nad}} \,=\,\l(\frac{c_{a}^2 - c_{e}^2}{4 \pi G a^2}\r)k^2\Phi + \Delta,\label{Eqn: delta p nad multi field/fluid}
\eeq
where
\beq
\Delta  \,=\,\sum_{i,j=1}^{n}\l[a_{ij}\delta s_{ij} \,+\, b_{ij}\delta\sigma_{ij}\r].\label{Eqn: Delta definition}
\eeq
The quantities $\delta s_{ij}$ and $\delta\sigma_{ij}$ are defined in Eqs.~(\ref{Eqn: delta s ij}) and (\ref{Eqn: delta sigma ij}), while the background-dependent coefficients $a_{ij}$ and $b_{ij}$ are defined in Eqs.~(\ref{Eqn: aij multi field/fluid}) and (\ref{Eqn: bij multi field/fluid}), respectively.

Substituting Eq.~(\ref{Eqn: delta p nad multi field/fluid}) into Eq.~(\ref{Eqn: Curvature Perturbation R dot-1}), it follows that:
\beq
\dot{\mathcal{R}} = -\l(\frac{H\,c_{e}^{2}}{\bar{\rho} + \bar{p}}\r)\l(\frac{k^2\Phi}{4 \pi G a^{2}}\r) \,+\,\frac{H\,\Delta}{\bar{\rho} + \bar{p}}. \label{Eqn: Curvature Perturbation R dot-2}
\eeq
In the above expression, $c_{e}$ is the total effective speed of sound defined in Eq.~(\ref{Eqn: ce2 total in multi field/fluid}). For a system consisting of a single matter component, whether it be a barotropic perfect fluid or a (canonical or non-canonical) scalar field, the entropy term $\Delta$ vanishes. Consequently, Eq.~(\ref{Eqn: Curvature Perturbation R dot-2}) implies that the curvature perturbation $\mathcal{R}$ is conserved on super-horizon scales ($k \rightarrow 0$) when the dynamics of the universe are driven by a single component.
However, this conservation law does not necessarily hold in a multi-field/fluid universe where the isocurvature perturbation term $\Delta$ is non-zero. In such scenarios, the non-vanishing $\Delta$ acts as a source for the evolution of the curvature perturbation $\mathcal{R}$, even on super-horizon scales.

From Eq.~(\ref{Eqn: Curvature Perturbation R dot-2}), the second-order evolution equation for the curvature perturbation in conformal time is found to be:
\beq
\mathcal{R}'' + 2\l(\frac{z'}{z}\r)\mathcal{R}' + c_{e}^{2}k^2\mathcal{R} \,=\, \frac{a^{4}(1+\epsilon)\Delta}{z^{2}c_{e}^{2}} \,+\, \l(\frac{a}{z^{2}H}\r)\l[\frac{a^{2}\Delta}{c_{e}^{2}}\r]', \label{Eqn: Curvature Perturbation R dot-dot}
\eeq
where the prime ($'$) denotes a derivative with respect to the conformal time $\eta$, defined by $\mathrm{d}\eta = \mathrm{d}t/a(t)$. In the above equation, the variable $z$ and the slow-roll parameter $\epsilon$ are defined as:
\ber
z &=& \frac{a\sqrt{\bar{\rho} + \bar{p}}}{H c_{e}}, \\
\epsilon &=& -\frac{\dot{H}}{H^{2}}.
\eer

The evolution of the curvature perturbation, as described by Eq.~(\ref{Eqn: Curvature Perturbation R dot-dot}), is valid for any system comprising $n$ minimally coupled canonical scalar fields, non-canonical scalar fields, and barotropic perfect fluids. It is evident from Eq.~(\ref{Eqn: Curvature Perturbation R dot-dot}) that in the large-$k$ limit, the equation reduces to $\mathcal{R}'' \approx -c_{e}^{2}k^2\mathcal{R}$. Consequently, in such a multi-field/fluid system, the curvature perturbation $\mathcal{R}$ propagates with the effective speed of sound $c_{e}$, defined in Eq.~(\ref{Eqn: ce2 total in multi field/fluid}), at small scales, i.e., at scales where $c_{e}k \gg aH$.

\section{Closed set of equations for perturbations in a multi-field/fluid system}\label{Sec: Closed set of equation}

Considering a universe where the dynamics are driven by a system of $n$ minimally coupled matter components, comprising barotropic perfect fluids, canonical scalar fields, and non-canonical scalar fields, we now derive a closed set of equations. These equations govern the evolution of both the background and the perturbations within the framework of the line element defined in Eq.~(\ref{Eqn: perturbed FRW line element}).

In the background homogeneous and isotropic universe described by the line element in Eq.~(\ref{Eqn: FRW line element}), the evolution of the scale factor $a(t)$ is governed by the set of equations (\ref{Eqn: Friedmann eqn multifield}) and (\ref{Eqn: Conservation Eqn multifield}). The conservation equation, $\dot{\bar{\rho}}_{_i} = -3H(\bar{\rho}_{_i} + \bar{p}_{_i})$, is satisfied by each matter component independently. To solve this equation, the relationship between $\bar{p}_{_i}$ and $\bar{\rho}_{_i}$ must be specified. For barotropic perfect fluids, one often expresses this as $\bar{p}_{_i} = w_i\bar{\rho}_{_i}$, where $w_i$ is the equation of state parameter characterizing the fluid. Given $w_i$, the conservation equation, expressed equivalently as $\mathrm{d}(\ln\bar{\rho}_{_i}) = -3(1 + w_i)\mathrm{d}(\ln a)$, can be integrated to determine the evolution of $\bar{\rho}_{_i}$ with respect to the scale factor $a$. 

Similarly, for a general barotropic perfect fluid defined by $\bar{p}_{_i} = f(\bar{\rho}_{_i})$, the conservation equation $\mathrm{d}\bar{\rho}_{_i} = -3\l[\bar{\rho}_{_i} + f(\bar{\rho}_{_i})\r]\mathrm{d}(\ln a)$ determines the functional form of $\bar{\rho}_{_i}(a)$. These solutions for $\bar{\rho}_{_i}(a)$ are then substituted into the Friedmann equation~(\ref{Eqn: Friedmann eqn multifield}) to solve for the time evolution of the scale factor $a(t)$.

In the case where the matter component is described by a generic scalar field with Lagrangian $\mathcal{L}_{i}(X_i,\phi_i)$, the energy density $\bar{\rho}_{_i}$ and pressure $\bar{p}_{_i}$ are given by Eqs.~(\ref{Eqn: density non canonical scalar field}) and (\ref{Eqn: pressure non canonical scalar field}), respectively. Consequently, the equation of state parameter $w_i = \bar{p}_{_i}/\bar{\rho}_{_i}$ is:
\beq
w_i = \frac{\mathcal{L}_i}{2X_i\mathcal{L}_{X_i} - \mathcal{L}_i}.\label{Eqn: w phi}
\eeq
Generally, we do not determine the evolution of $\bar{\rho}_{_i}$ as a function of the scale factor $a$ by solving the conservation equation $\mathrm{d}(\ln\bar{\rho}_{_i}) = -3(1 + w_i)\mathrm{d}(\ln a)$ directly via $w_i$. Instead, we solve the Klein-Gordon equation for the scalar field to determine $\bar{\phi}_{_i}(t)$ and subsequently find the evolution of $\bar{\rho}_{_i}(t)$ using Eq.~(\ref{Eqn: density non canonical scalar field}). The field equation for $\bar{\phi}_{_i}$ follows from the Euler-Lagrange equation:
\beqstr
\frac{\partial \mathcal{L}_i}{\partial \phi_{_i}} - \l(\frac{\partial \mathcal{L}_i}{\partial (\partial_\mu \phi_{_i})}\r)_{;\,\mu}\,=\,0,
\eeqstr
which is equivalent to the conservation equation $\dot{\bar{\rho}}_{_i} = -3H(\bar{\rho}_{_i} + \bar{p}_{_i})$. For a non-canonical field, this leads to the following evolution equation:
\beq
\l(\mathcal{L}_{X_i} + 2X\mathcal{L}_{X_i X_i} \r)\ddot{\bar{\phi}}_{_i}  + 3H\dot{\bar{\phi}}\mathcal{L}_{X_i}  + 2X\mathcal{L}_{X_i \phi_i} - \mathcal{L}_{\phi_i}\,=\,0.\label{Eqn: Field eqn for phi NC field}
\eeq
For each component described by a scalar field, solving the above yields $\bar{\phi}_i (t)$ and $\bar{\rho}_i (t)$, which are then substituted into the Friedmann equation~(\ref{Eqn: Friedmann eqn multifield}). Thus, for a multi-field/fluid system, the Friedmann equation together with the conservation equations for the barotropic perfect fluids and the field equations for the scalar fields form a closed set of equations governing the evolution of the background universe.

To determine the evolution of perturbations about the homogeneous and isotropic background universe containing $n$ minimally coupled fields and fluids, only a single variable is required to describe the metric perturbations. As discussed in Sec.~\ref{Sec: Basic eqn CPT}, out of the four variables describing the scalar perturbations in the line element~(\ref{Eqn: perturbed FRW line element}), two can be eliminated through the gauge transformations defined in Eqs.~(\ref{Eqn: gauge transformation phi}) through (\ref{Eqn: gauge transformation E}). Equivalently, the scalar degrees of freedom in the metric can be described by the two gauge-invariant Bardeen potentials, $\Phi$ and $\Psi$, defined in Eqs.~(\ref{Eqn: Bardeen Potential Phi}) and (\ref{Eqn: Bardeen Potential Psi}). However, for matter components such as barotropic perfect fluids and canonical or non-canonical scalar fields, there is no anisotropic stress; that is, $\delta T^{i}_{\hspace{0.2cm}j} = 0$ for all $i \neq j$ ($i,j = 1,2,3$). As noted in Sec.~\ref{Sec: Basic eqn CPT}, this absence of anisotropic stress ensures the equality of the two Bardeen potentials ($\Phi = \Psi$). Consequently, only one variable, $\Phi$, is necessary to describe the scalar degree of freedom of the metric perturbations.

For a universe driven by a single barotropic perfect fluid or a single scalar field, the evolution of the Bardeen potential $\Phi$ is governed by Eq.~(\ref{Eqn: SF Phi dot dot Eqn}). Thus, a single second-order differential equation is sufficient to describe the perturbations of the system. Once the evolution of $\Phi$ is determined, the evolution of the gauge-invariant quantities $\delta \rho^{^{(gi)}}_{_B}$, $\delta q^{^{(gi)}}_{_B}$, and $\delta p^{^{(gi)}}_{_B}$ can be directly obtained using the Einstein field equations, specifically Eqs.~(\ref{Eqn: GI Perturbed Einstein Eqn-1}), (\ref{Eqn: GI Perturbed Einstein Eqn-2}), and (\ref{Eqn: GI Perturbed Einstein Eqn-3}).

In the case of a multi-field/fluid system, the evolution of $\Phi$ is governed by Eq.~(\ref{Eqn: multi field Phi dot dot Eqn}). Notably, the gauge-invariant isocurvature terms $\delta s_{ij}$ and $\delta\sigma_{ij}$, defined in Eqs.~(\ref{Eqn: delta s ij}) and (\ref{Eqn: delta sigma ij}), respectively, act as source terms on the right-hand side of the evolution equation for $\Phi$. To obtain a closed set of equations, one must also include the differential equations governing the evolution of $\delta s_{ij}$ and $\delta\sigma_{ij}$. These isocurvature perturbations can be expressed in terms of the individual component perturbations as:
\ber
\delta s_{ij} &=& (\bar{\rho} + \bar{p})\l[\frac{\delta \rho^{^{(gi)}}_{_{Bi}}}{\bar{\rho}_{_i} + \bar{p}_{_i}}\,-\,\frac{\delta \rho^{^{(gi)}}_{_{Bj}}}{\bar{\rho}_{_j} + \bar{p}_{_j}}\r],\label{Eqn: delta s ij bardeen delta rho}\\
\delta\sigma_{ij} \,&=&\, 3H(\bar{\rho} + \bar{p})\l[\frac{\delta q^{^{(gi)}}_{_{Bi}}}{\bar{\rho}_{_i} + \bar{p}_{_i}}\,-\,\frac{\delta q^{^{(gi)}}_{_{Bj}}}{\bar{\rho}_{_j} + \bar{p}_{_j}}\r], \label{Eqn: delta sigma ij bardeen delta rho}
\eer
where $\delta \rho^{^{(gi)}}_{_B}$ and $\delta q^{^{(gi)}}_{_B}$ are the gauge-invariant Bardeen density and momentum perturbations defined in Eqs.~(\ref{Eqn: gauge invarian delta rho}) and (\ref{Eqn: gauge invarian delta q}). By utilizing the evolution equations for $\delta \rho^{^{(gi)}}_{_B}$ and $\delta q^{^{(gi)}}_{_B}$ from Eqs.~(\ref{Eqn: GI delta rho conservation eqn}) and (\ref{Eqn: GI delta q conservation eqn}), we find:
\ber
\dot{\delta s_{ij}} &=& -3H(1 + c_{a}^2)\delta s_{ij} + \l(\frac{k^{2}}{3H^{2}a^{2}}\r)(H\delta\sigma_{ij})
 + \l(\frac{3H\l[(c_{ei}^2- c_{ai}^2) - (c_{ej}^2-c_{aj}^2)\r]}{4\pi Ga^{2}}\r)k^{2}\Phi\hspace{1cm} \nonumber \\ 
&& - \l(\frac{3H}{\bar{\rho} + \bar{p}}\r) \sum_{l = 1}^{n}(\bar{\rho}_{_l} + \bar{p}_{_l})\l[(c_{ei}^2- c_{ai}^2)(\delta s_{il} - \delta\sigma_{il})
 -  (c_{ej}^2- c_{aj}^2)(\delta s_{jl} - \delta\sigma_{jl})\r],\label{Eqn: del s-ij eqn}
\eer 
\ber
\dot{\delta\sigma_{ij}} &=& -3H(1 + c_{a}^2)\delta\sigma_{ij} + \l(\frac{\dot{H}}{H}\r)\delta\sigma_{ij}
 + \l(\frac{3H(c_{ei}^2- c_{ej}^2)}{4\pi Ga^{2}}\r)k^{2}\Phi \hspace{1cm}\nonumber \\
  && - \l(\frac{3H}{\bar{\rho} + \bar{p}}\r)\sum_{l = 1}^{n}(\bar{\rho}_{_l} + \bar{p}_{_l})\l[(c_{ei}^2\delta s_{il} - c_{ej}^2\delta s_{jl}) -(c_{ei}^2\delta \sigma_{il} - c_{ej}^2\delta \sigma_{jl})\r].\hspace{0.5cm}\label{Eqn: del sigma-ij eqn}
\eer
These two equations describe the dynamics of the isocurvature sources that drive the evolution of the Bardeen potential $\Phi$ in Eq.~(\ref{Eqn: multi field Phi dot dot Eqn}).

From the definitions of $\delta s_{ij}$ and $\delta\sigma_{ij}$ for $i, j = 1, \dots, n$ in Eqs.~(\ref{Eqn: delta s ij}) and (\ref{Eqn: delta sigma ij}), or equivalently Eqs.~(\ref{Eqn: delta s ij bardeen delta rho}) and (\ref{Eqn: delta sigma ij bardeen delta rho}), it is clear that both $\delta s_{ij}$ and $\delta\sigma_{ij}$ are anti-symmetric. Consequently, each possesses $n(n-1)/2$ non-zero components. However, not all of these $n(n-1)/2$ terms are independent. In fact, we can express any $\delta s_{ij}$ and $\delta\sigma_{ij}$ as:
\ber
\delta s_{ij} &=& \delta s_{1j} - \delta s_{1i},\label{Eqn: delta s ij in terms of s-1i}\\
\delta\sigma_{ij} &=& \delta\sigma_{1j} - \delta\sigma_{1i}.\label{Eqn: delta sigma ij in terms of s-1i}
\eer
Therefore, the $n(n-1)/2$ non-zero terms in $\delta s_{ij}$ are fully determined by the $(n-1)$ quantities $\{\delta s_{12}, \delta s_{13}, \dots, \delta s_{1n}\}$. Similarly, the $n(n-1)/2$ non-zero terms in $\delta \sigma_{ij}$ are derived from the $(n-1)$ quantities $\{\delta \sigma_{12}, \delta \sigma_{13}, \dots, \delta \sigma_{1n}\}$.

By substituting Eqs.~(\ref{Eqn: delta s ij in terms of s-1i}) and (\ref{Eqn: delta sigma ij in terms of s-1i}) into Eq.~(\ref{Eqn: relation delta rho p and q multi filed/fluid}), we can also express the total pressure perturbation $\delta p$ as:
\beq
\delta p \,= c_{e}^{2}\,\delta \rho +  3H\l(c_{a}^{2} - c_{e}^2\r) \delta q + \sum_{j=2}^{n}\l[A_{j}\delta s_{1j} \,+\, B_{j}\delta\sigma_{1j}\r], \label{Eqn: new relation delta rho p and q multi field/fluid}
\eeq
where the coefficients $A_j$ and $B_j$ are given by:
\ber
A_{j} &=& \l(\frac{\bar{\rho}_{_j} + \bar{p}_{_j}}{\bar{\rho}+ \bar{p}}\r)\l[c_{e}^{2}\,-\,c_{ej}^{2}\r], \\
B_{j} &=& \l(\frac{\bar{\rho}_{_j} + \bar{p}_{_j}}{\bar{\rho}+ \bar{p}}\r)\l[(c_{a}^{2} - c_{e}^{2})\,-\, (c_{aj}^{2} - c_{ej}^{2})\r].
\eer
Consequently, the evolution of $\Phi$ in Eq.~(\ref{Eqn: multi field Phi dot dot Eqn}) can be rewritten as:
\ber
\ddot{\Phi} + H\l[4 + 3c_a^2\r]\dot{\Phi} + \l[2\dot{H} + 3H^{2}(1 + c_a^2)\r]\Phi 
+ \l(\f{c_e^2 k^2}{a^2}\r)\Phi
 = 4 \pi G \sum_{j=2}^{n}\l[A_{j}\delta s_{1j} + B_{j}\delta\sigma_{1j}\r]. \label{Eqn: new multi field Phi dot dot Eqn}
\eer
The evolution of the source terms $\delta s_{1j}$ and $\delta\sigma_{1j}$ for $j = 2, \dots, n$ are governed by:
\ber
\dot{\delta s_{1j}} &=& -3H(1 + c_{a}^2)\delta s_{1j}- 3H(c_{ej}^2- c_{aj}^2)(\delta s_{1j} - \delta\sigma_{1j})
+ \frac{k^{2}\delta \sigma_{1j}}{3Ha^{2}}\nonumber\\
 && + 3H\l[(c_{e1}^2- c_{a1}^2) - (c_{ej}^2-c_{aj}^2)\r]\l[\frac{k^{2}\Phi}{4\pi Ga^{2}} - \sum_{l = 2}^{n}\l(\frac{\bar{\rho}_{_l} + \bar{p}_{_l}}{\bar{\rho} + \bar{p}}\r)(\delta s_{1l} - \delta\sigma_{1l})\r] ~~~~ \label{Eqn: del s-1j eqn}\\
\dot{\delta\sigma_{1j}} &=& -3H(1 + c_{a}^2)\delta\sigma_{1j}   + \l(\frac{3H(c_{e1}^2- c_{ej}^2)}{4\pi Ga^{2}}\r)k^{2}\Phi
+ \l(\frac{\dot{H}}{H}\r)\delta\sigma_{1j} \nonumber\\
&& - \l(\frac{3H}{\bar{\rho} + \bar{p}}\r)\sum_{l = 1}^{n}(\bar{\rho}_{_l} + \bar{p}_{_l})\l[(c_{e1}^2\delta s_{1l} - c_{ej}^2\delta s_{jl}) -(c_{e1}^2\delta \sigma_{1l} - c_{ej}^2\delta \sigma_{jl})\r]. \label{Eqn: del sigma-1j eqn}
\eer
The $(2n-1)$ equations provided in Eqs.~(\ref{Eqn: new multi field Phi dot dot Eqn}), (\ref{Eqn: del s-1j eqn}), and (\ref{Eqn: del sigma-1j eqn}) constitute a closed system for the $(2n-1)$ variables: $\Phi$, $\{\delta s_{1j}\}$, and $\{\delta \sigma_{1j}\}$. 

In a universe driven by $n$ minimally coupled components, there are initially $(3n+1)$ variables describing the perturbations: $(i)$ the gauge-invariant Bardeen potential $\Phi$ for the metric, and $(ii)$ the set $\{\delta \rho^{^{(gi)}}_{_{Bi}}, \delta p^{^{(gi)}}_{_{Bi}}, \delta q^{^{(gi)}}_{_{Bi}}\}$ for $i = 1, \dots, n$, describing the energy-momentum tensor perturbations of each component. However, since the $\delta p^{^{(gi)}}_{_{Bi}}$ are related to $\delta \rho^{^{(gi)}}_{_{Bi}}$ and $\delta q^{^{(gi)}}_{_{Bi}}$ via Eq.~(\ref{Eqn: multiF relation delta rho pi and qi}), these $n$ relations reduce the independent quantities to $(2n+1)$. Furthermore, given that $\delta \rho^{^{(gi)}}_{_{B}} = \sum \delta \rho^{^{(gi)}}_{_{Bi}}$ and $\delta q^{^{(gi)}}_{_{B}} = \sum \delta q^{^{(gi)}}_{_{Bi}}$, the Einstein equations (\ref{Eqn: GI Perturbed Einstein Eqn-1}) and (\ref{Eqn: GI Perturbed Einstein Eqn-2}) act as two additional constraints, further reducing the independent functions to $(2n-1)$. 
While this $(2n-1)$ set provides a complete description of the system, it is important to note that Eq.~(\ref{Eqn: new multi field Phi dot dot Eqn}) is a second-order equation. Thus, the system effectively possesses $2n$ degrees of freedom, requiring $2n$ initial conditions, namely the values of $\Phi, \dot{\Phi}, \delta s_{1j}$, and $\delta \sigma_{1j}$ are required to be specified at an initial time $t = t_i$.

Once the $(2n-1)$ quantities, i.e., $\Phi(t)$, $\delta s_{1j}(t)$, and $\delta \sigma_{1j}(t)$, are determined for a given wave number $k$ using the closed system of equations (\ref{Eqn: new multi field Phi dot dot Eqn}), (\ref{Eqn: del s-1j eqn}), and (\ref{Eqn: del sigma-1j eqn}), the gauge-invariant component density perturbations $\delta \rho^{^{(gi)}}_{_{Bi}}$ can be reconstructed. This is achieved by combining the Poisson equation (\ref{Eqn: GI Perturbed Einstein Eqn-1}) with the definition of the isocurvature perturbation (\ref{Eqn: delta s ij bardeen delta rho}), noting that the total density perturbation is the sum of the individual components, $\delta \rho^{^{(gi)}}_{_B} = \sum_{i=1}^{n}\delta \rho^{^{(gi)}}_{_{Bi}}$. The resulting expressions for $\delta \rho^{^{(gi)}}_{_{Bi}}$ in terms of $\Phi$ and $\delta s_{ij}$ are:
\ber
\delta \rho^{^{(gi)}}_{_{B1}} &=& \l(\frac{\bar{\rho}_{_1} + \bar{p}_{_1}}{\bar{\rho}+ \bar{p}}\r)\sum_{j=2}^{n}\l(\frac{\bar{\rho}_{_j} + \bar{p}_{_j}}{\bar{\rho}+ \bar{p}}\r)\delta s_{1j} 
- \l(\frac{\bar{\rho}_{_1} + \bar{p}_{_1}}{4\pi G (\bar{\rho}+ \bar{p})}\r)\l(3H^{2}\Phi + 3H\dot{\Phi} + \frac{k^{2}}{a^{2}}\Phi \r),\label{Eqn: delta rho-1 from closed set}\\
\delta \rho^{^{(gi)}}_{_{Bj}} &=& \l(\bar{\rho}_{_j} + \bar{p}_{_j}\r)\l[\frac{\delta \rho^{^{(gi)}}_{_{B1}}}{\bar{\rho}_{_1} + \bar{p}_{_1}} \;-\; \frac{\delta s_{1j}}{\bar{\rho}+ \bar{p}}\r],\label{Eqn: delta rho-j from closed set}
\eer
where $j = 2, \dots, n$. Similarly, the gauge-invariant momentum density perturbations $\delta q^{^{(gi)}}_{_{Bj}}$ can be evaluated using the momentum constraint in Eq.~(\ref{Eqn: GI Perturbed Einstein Eqn-2}) and the isocurvature velocity definition in Eq.~(\ref{Eqn: delta sigma ij bardeen delta rho}). The components $\delta q^{^{(gi)}}_{_{Bi}}$ are given by:
\ber
\delta q^{^{(gi)}}_{_{B1}} &=& \l(\frac{\bar{\rho}_{_1} + \bar{p}_{_1}}{\bar{\rho}+ \bar{p}}\r)\sum_{j=2}^{n}\,\l(\frac{\bar{\rho}_{_j} + \bar{p}_{_j}}{3H(\bar{\rho}+ \bar{p})}\r)\delta\sigma_{1j} - \l(\frac{\bar{\rho}_{_1} + \bar{p}_{_1}}{4\pi G(\bar{\rho}+ \bar{p})}\r)\l(\dot{\Phi} + H\Phi \r),\label{Eqn: delta q-1 from closed set}\\
\delta q^{^{(gi)}}_{_{Bj}} &=& \l(\bar{\rho}_{_j} + \bar{p}_{_j}\r)\l[\frac{\delta q^{^{(gi)}}_{_{B1}}}{\bar{\rho}_{_1} + \bar{p}_{_1}} \;-\; \frac{\delta\sigma_{1j}}{3H(\bar{\rho}+ \bar{p})}\r],\label{Eqn: delta q-j from closed set}
\eer
for $j = 2, \dots, n$. Furthermore, once all components of $\delta \rho^{^{(gi)}}_{_{Bi}}$ and $\delta q^{^{(gi)}}_{_{Bi}}$ are determined, the corresponding gauge-invariant pressure perturbations $\delta p^{^{(gi)}}_{_{Bi}}$ for each component can be calculated using Eq.~(\ref{Eqn: multiF relation delta rho pi and qi}).

Note that in the longitudinal gauge, defined as the gauge where $E = B = 0$ in the perturbed line element~(\ref{Eqn: perturbed FRW line element}), one obtains the identical closed set of equations given in Eqs.~(\ref{Eqn: new multi field Phi dot dot Eqn}) through (\ref{Eqn: del sigma-1j eqn}), along with the corresponding expressions for $\delta \rho$ and $\delta q$ in Eqs.~(\ref{Eqn: delta rho-1 from closed set}) through (\ref{Eqn: delta q-j from closed set}). This equivalence arises because all gauge-invariant Bardeen variables, such as $\Phi$, $\delta \rho^{^{(gi)}}_{_{B}}$, $\delta p^{^{(gi)}}_{_{B}}$, and $\delta q^{^{(gi)}}_{_{B}}$, coincide with the actual physical perturbations evaluated in the longitudinal gauge. This is directly evident from the definitions in Eqs.~(\ref{Eqn: gauge invarian delta rho}) through (\ref{Eqn: gauge invarian delta q}), as well as the potential definitions in Eqs.~(\ref{Eqn: Bardeen Potential Phi}), (\ref{Eqn: Bardeen Potential Psi}), and the constraint in Eq.~(\ref{Eqn: equality of Bardeen Potentials}).

An alternative method for expressing a closed set of equations is to utilize the energy-momentum conservation equations for $\delta \rho^{^{(gi)}}_{_{Bi}}$ and $\delta q^{^{(gi)}}_{_{Bi}}$, as given in Eqs.~(\ref{Eqn: GI delta rho conservation eqn}) and (\ref{Eqn: GI delta q conservation eqn}), for each matter component $i = 1, \dots, n-1$, alongside Eq.~(\ref{Eqn: GI Perturbed Einstein Eqn-3}) for the evolution of $\Phi$. In Eq.~(\ref{Eqn: GI Perturbed Einstein Eqn-3}), the total gauge-invariant pressure perturbation is the sum over all components, $\delta p^{^{(gi)}}_{_{B}} = \sum_{i = 1}^{n}\delta p^{^{(gi)}}_{_{Bi}}$, where each $\delta p^{^{(gi)}}_{_{Bi}}$ is determined by Eq.~(\ref{Eqn: multiF relation delta rho pi and qi}). Consequently, Eq.~(\ref{Eqn: GI Perturbed Einstein Eqn-3}) and the $2(n-1)$ conservation equations for the first $n-1$ components form a closed system of $(2n-1)$ equations. 
Since Eq.~(\ref{Eqn: GI Perturbed Einstein Eqn-3}) is of second order, this formulation likewise confirms the existence of $2n$ degrees of freedom in the system. Once this closed set of equations is solved, the remaining perturbations for the $n^{\mathrm{th}}$ component, specifically $\delta \rho^{^{(gi)}}_{_{Bn}}$ and $\delta q^{^{(gi)}}_{_{Bn}}$, can be reconstructed using the two constraint equations provided by the Poisson equation (\ref{Eqn: GI Perturbed Einstein Eqn-1}) and the momentum constraint (\ref{Eqn: GI Perturbed Einstein Eqn-2}), respectively.

Alternatively, one can also describe the evolution of perturbations using a set of $2n$ first-order differential equations. In this approach, the evolution of the Bardeen potential $\Phi$ is determined by the first-order Einstein equation, Eq.~(\ref{Eqn: GI Perturbed Einstein Eqn-1}), where the total density perturbation is given by $\delta \rho^{^{(gi)}}_{_B} = \sum_{i = 1}^{n}\delta \rho^{^{(gi)}}_{_{Bi}}$. The dynamics of each component density perturbation, $\delta \rho^{^{(gi)}}_{_{Bi}}$, are governed by the conservation equation, Eq.~(\ref{Eqn: GI delta rho conservation eqn}), for $i = 1, \dots, n$. Furthermore, the momentum density perturbations $\delta q^{^{(gi)}}_{_{Bj}}$ are evolved using Eq.~(\ref{Eqn: GI delta q conservation eqn}) for $j = 1, \dots, n-1$. This formulation results in a system of $2n$ first-order equations for the variables $\Phi$, $\{\delta \rho^{^{(gi)}}_{_{Bi}}\}$, and $\{\delta q^{^{(gi)}}_{_{Bj}}\}$. The remaining $n^{\mathrm{th}}$ momentum term, $\delta q^{^{(gi)}}_{_{Bn}}$, is then determined algebraically through the momentum constraint equation, Eq.~(\ref{Eqn: GI Perturbed Einstein Eqn-2}), in conjunction with Eq.~(\ref{Eqn: GI Perturbed Einstein Eqn-1}), yielding:
\ber
\delta q^{^{(gi)}}_{_{Bn}} &=& \l(\frac{1}{12\pi G Ha^2 }\r)k^{2}\Phi 
+ \l(\frac{1}{3 H }\r)\l[\sum\limits_{i = 1}^{n}\delta \rho^{^{(gi)}}_{_{Bi}}   - 3 H \sum\limits_{j = 1}^{n-1}\delta q^{^{(gi)}}_{_{Bj}}\r]. 
\eer

It is evident that multiple equivalent formulations exist for expressing a closed set of equations to describe the evolution of perturbations in a multi-field/fluid universe. However, as demonstrated in the subsequent section, the representation provided by Eqs.~(\ref{Eqn: new multi field Phi dot dot Eqn}) through (\ref{Eqn: del sigma-1j eqn}) offers a distinct advantage. By explicitly treating the isocurvature perturbation terms, $\delta s_{ij}$ and $\delta \sigma_{ij}$, as source terms in the evolution equation for the Bardeen potential $\Phi$, this approach provides a clear physical mapping of how entropy modes drive metric fluctuations. Notably, these isocurvature terms vanish in a universe dominated by a single matter component or in scenarios where the perturbations are purely adiabatic, reducing the system to the standard single-field description.

\section{Large-scale behavior of perturbations in a multi-field/fluid cosmological system}\label{Sec: Large scale behavior}

In a multi-field/fluid universe comprising barotropic perfect fluids, canonical scalar fields, and non-canonical scalar fields, the dynamics of perturbations are governed by Eqs.~(\ref{Eqn: new multi field Phi dot dot Eqn}) through (\ref{Eqn: del sigma-1j eqn}). It is important to note that in Eq.~(\ref{Eqn: new multi field Phi dot dot Eqn}), $c_e$ represents the effective speed of sound of the composite system, as defined in Eq.~(\ref{Eqn: ce2 total in multi field/fluid}). In scenarios where the universe contains only a single matter component, the evolution of the Bardeen potential $\Phi$ is determined by Eq.~(\ref{Eqn: SF Phi dot dot Eqn}). However, in a multi-field/fluid system, the isocurvature perturbation terms, such as $\delta s_{ij}$ and $\delta\sigma_{ij}$, act as non-vanishing source terms in the evolution equation for $\Phi$.

Let us first consider a scenario where the universe consists of $n$ matter components, specifically barotropic perfect fluids and pure kinetic non-canonical scalar fields. For both types of matter, as discussed in Sec.~\ref{Sec: single fluid/field}, the effective speed of sound equals the adiabatic speed of sound, $c_{ei}^2 = c_{ai}^2$~\cite{Unnikrishnan:2010ag,Arroja:2010wy}. Consequently, the effective speed of sound for the total system is identically equal to the adiabatic speed of sound, i.e., $c_{e}^2 = c_{a}^2$. 

In the large-scale limit ($k \to 0$), Eq.~(\ref{Eqn: del s-1j eqn}) simplifies to:
\beq
\dot{\delta s_{ij}} + 3H(1 + c_{a}^2)\delta s_{ij} = 0.
\eeq
This implies that in the $k \to 0$ limit:
\beq
\delta s_{ij} \propto (\bar{\rho} + \bar{p}) \implies \l[\frac{\delta \rho^{^{(gi)}}_{_{Bi}}}{\bar{\rho}_{_i} + \bar{p}_{_i}}\,-\,\frac{\delta \rho^{^{(gi)}}_{_{Bj}}}{\bar{\rho}_{_j} + \bar{p}_{_j}}\r] = \text{constant}.
\eeq
Thus, the relative entropy perturbation between these components is conserved on super-horizon scales~\cite{Malik:2004tf}.
Furthermore, if the universe additionally contains canonical or general non-canonical scalar fields, Eq.~(\ref{Eqn: del s-ij eqn}) ensures that $\delta s_{ij}/(\bar{\rho} + \bar{p})$ remains conserved as $k \to 0$, provided that both the $i^{\mathrm{th}}$ and $j^{\mathrm{th}}$ components are either barotropic perfect fluids or pure kinetic non-canonical scalar fields.

In the general case of a multi-field/fluid universe, let us consider a scenario where the initial perturbations are purely adiabatic in the large-scale limit ($k \to 0$). Mathematically, this implies that the non-adiabatic pressure perturbation $\delta p_{\mathrm{nad}}$, defined in Eq.~(\ref{Eqn: delta p nad multi field/fluid}), is zero at the initial time $t_i$ for $k \to 0$. This condition is satisfied if the isocurvature terms vanish initially, i.e., $\delta s_{1j} = \delta \sigma_{1j} = 0$ for all $j = 2, \dots, n$ at $t = t_i$. 
From the evolution equations (\ref{Eqn: del s-1j eqn}) and (\ref{Eqn: del sigma-1j eqn}), it follows that if these isocurvature components are zero at $k = 0$ initially, they remain zero throughout the subsequent evolution of the universe. Consequently, as per Eq.~(\ref{Eqn: delta p nad multi field/fluid}), $\delta p_{\mathrm{nad}}$ remains zero at the $k = 0$ scale, ensuring that the perturbations remain adiabatic at all times on super-horizon scales.

Thus, we have demonstrated that in a multi-field/fluid universe comprising barotropic perfect fluids, canonical scalar fields, and non-canonical scalar fields, perturbations that are initially adiabatic at $k = 0$ remain adiabatic throughout the evolution of the universe. This extends the well-known result to a multi-field/fluid universe consisting of non-canonical scalar fields~\cite{Wands:2000dp,Gordon:2000hv,Weinberg:2003sw}. Consequently, since $\delta p_{\mathrm{nad}} = 0$ in this limit, the total pressure perturbation on super-horizon scales ($k \to 0$) is related to the total energy density perturbation via the adiabatic speed of sound:
\beq
\delta p = c_a^2 \delta \rho \quad \text{at } k = 0, \label{Eqn: adiabaticity at k = 0}
\eeq
where $c_a^2$ is the square of the adiabatic speed of sound for the composite system. 
It is crucial to note that this relationship does not imply that the physical speed of sound at $k = 0$ is $c_a$. The effective speed of sound of the total system remains uniquely defined by Eq.~(\ref{Eqn: ce2 total in multi field/fluid}). As established by the wave equations (\ref{Eqn: wave eqn Phi}) and (\ref{Eqn: wave eqn delta rho}), $c_e$ is the characteristic speed at which perturbations such as $\Phi$ propagate at small scales (the large-$k$ limit).

Interestingly, it follows from Eq.~(\ref{Eqn: adiabaticity at k = 0}) that a multi-field/fluid universe with initially adiabatic perturbations at the $k = 0$ scale is dynamically equivalent to a universe containing a single barotropic perfect fluid at that super-horizon $k \rightarrow 0$ scales.

\section{Implications for unified dark sector models}\label{Sec: Dark Sector models}

A wealth of cosmological observations, including Type Ia supernova light curves, Cosmic Microwave Background (CMB) anisotropies, Baryon Acoustic Oscillations (BAO), and Large Scale Structure (LSS) data, consistently indicate that the Universe is currently undergoing a phase of accelerated expansion~\cite{Planck:2018vyg,SupernovaSearchTeam:1998fmf,SupernovaCosmologyProject:1998vns,DESI:2024mwx}. According to the second Friedmann equation~(\ref{Eqn: Friedman Eqn-2}), this acceleration ($\ddot{a} > 0$) requires the dominant matter component of the present-day Universe to satisfy the condition $\rho + 3p < 0$. This component, characterized by negative pressure, is referred to as dark energy. 
Current observational data suggest that dark energy constitutes approximately $70\%$ of the total energy density, with an equation of state parameter $w$ close to $-1$~\cite{DESI:2024mwx}. The remaining $30\%$ is primarily composed of pressureless dark matter. Despite their distinct physical properties, the energy densities of dark matter and dark energy are of the same order of magnitude at the present epoch. This ``coincidence" suggests that these two entities might be different manifestations of a single underlying substance. Consequently, unified dark sector models, often based on a single scalar field that mimics both dark matter and dark energy, have been extensively explored in the literature~\cite{Scherrer:2004au,Bertacca:2007ux,Bertacca:2010ct,Mishra:2018tki,Sahni:2015hbf}. In this section, we apply our formalism to analyze the evolution of the effective speed of sound in unified dark sector models driven by a single non-canonical scalar field.

Consider a scenario where a single non-canonical scalar field, characterized by the Lagrangian $L(X,\phi)$, mimics the dynamics of a universe containing dark matter and dark energy as distinct entities. For such a unified dark sector model to be phenomenologically viable, it must satisfy two primary conditions: the evolution of the scale factor $a(t)$ (background dynamics) and the evolution of the Bardeen potential $\Phi(t)$ (linear perturbations) must be identical to those in a universe containing dark matter and dark energy as independent, minimally coupled components.

The evolution of the scale factor $a(t)$ is governed by the Friedmann equation~(\ref{Eqn: Friedman Eqn-1}) together with the background conservation equations~(\ref{Eqn: background conservation eqn matter content 1}) and (\ref{Eqn: background conservation eqn matter content 2}) for the energy densities and pressures of dark matter and dark energy, respectively. Let $w_{\phi}$ be the equation of state parameter of the scalar field with Lagrangian $L(X,\phi)$ describing the unified dark sector, as defined in Eq.~(\ref{Eqn: w phi}). Since dark matter is treated as a pressureless barotropic fluid, its equation of state is $w_{m} = 0$. Let $w_{de}$ represent the equation of state for dark energy.

For a single scalar field to mimic the background expansion of both components, its total equation of state parameter $w_{\phi}$ must satisfy:
\beq
w_{\phi} = w_{de} \Omega_{de}, \label{Eqn: w-phi of unified dark model}
\eeq
where $\Omega_{de} = (8 \pi G \rho_{de})/(3H^{2})$ is the density parameter of the dark energy. Assuming $-1 < w_{de} < -1/3$, it follows that during the matter-dominated epoch when $\Omega_{de} \approx 0$, we have $w_{\phi} \approx 0$, effectively mimicking cold dark matter. However, at the present epoch, with $\Omega_{de} \approx 0.7$ and $w_{de} \approx -1$, Eq.~(\ref{Eqn: w-phi of unified dark model}) yields $w_{\phi} \approx -0.7$. Thus, for a single scalar field to successfully unify the dark sector, its equation of state $w_{\phi}$ must evolve from an initial value of zero to approximately $-0.7$ today. It is possible to reconstruct a specific form of the Lagrangian $L(X,\phi)$ to achieve this specific trajectory. When $w_{\phi}$ satisfies Eq.~(\ref{Eqn: w-phi of unified dark model}), the resulting evolution of the scale factor $a(t)$ is identical to that of a universe containing dark matter and dark energy as separate, independent entities.

Similarly, for the evolution of perturbations in the unified dark sector to mimic a universe where dark matter and dark energy are separate entities, the evolution of the Bardeen potential $\Phi$ must be identical in both cases. When the universe is driven by a single scalar field with Lagrangian $L(X,\phi)$, the dynamics of $\Phi$ are governed by Eq.~(\ref{Eqn: SF Phi dot dot Eqn}). In contrast, for a two-component field/fluid system, the evolution is governed by Eq.~(\ref{Eqn: TF Phi dot dot Eqn}).
In these governing equations, the effective speed of sound $c_e^2$ is defined by Eq.~(\ref{Eqn: c_e^2 for NC scalar field}) for the non-canonical scalar field and Eq.~(\ref{Eqn: ce2 effective speed of sound two field/fluid}) for the two-component system. Assuming the background evolution $a(t)$ is matched across both models, ensuring that $\Phi(t)$ remains identical for every wave number $k$ necessitates two conditions: (i) the contributions of the isocurvature terms, $\delta s_{ij}$ and $\delta \sigma_{ij}$, to the evolution of $\Phi$ must be negligible, and (ii) the effective speed of sound $c_e$ must be identical in both scenarios. Even if isocurvature effects are assumed to be negligible, the latter condition remains a necessary requirement for the equivalence of the two models.

Let $c_{e}^{2}$ represent the effective speed of sound for the unified dark sector model based on a single non-canonical scalar field. To achieve perturbative equivalence, this value must equal the total effective speed of sound of a universe where dark matter and dark energy are separate entities. Assuming dark matter is a pressureless barotropic fluid ($c_{e(m)}^{2} = w_{m} = 0$) and letting $c_{e(de)}^{2}$ be the effective speed of sound of the equivalent dark energy component, Eq.~(\ref{Eqn: ce2 effective speed of sound two field/fluid}) yields:
\beq
c_{e(de)}^{2} = c_{e}^{2}\l[ 1 + \frac{\Omega_{m}}{(1+w_{de})\Omega_{de}} \r],
\label{Eqn: ce2 of equivalent DE}
\eeq
where $\Omega_{m}$ is the dark matter density parameter. Assuming the dark energy component satisfies the null energy condition ($1+w_{de} > 0$), Eq.~(\ref{Eqn: ce2 of equivalent DE}) reveals a significant challenge: if a single canonical scalar field ($c_{e}^{2} = 1$) is used to mimic both components, the equivalent dark energy speed of sound becomes superluminous ($c_{e(de)}^{2} > 1$). 
This superluminality issue persists in unified models where $c_{e}^{2}$ remains constant, such as the purely kinetic or power-law models where $L(X,\phi) = X^{\alpha} - V(\phi)$. In these cases, even if $c_{e}^{2} < 1$, the factor $\Omega_{m} / [(1+w_{de})\Omega_{de}]$ grows significantly during the matter-dominated epoch, inevitably leading to an era where $c_{e(de)}^{2} > 1$. However, this problem is circumvented in models where $L(X,\phi) = X^{\alpha} - V_0$ with a constant potential $V_0$~\cite{Sahni:2015hbf}. In such models, the dark energy mimics a cosmological constant ($p_{de} = -\rho_{de}$) and $c_{{e(m)}} \neq 0$. Consequently, the total speed of sound reduces to that of the dark matter-like component, $c_{e}^{2} = c_{e(m)}^{2}$, ensuring the perturbations remain subluminal throughout the evolution.

During the matter-dominated epoch, for instance at a redshift of $z \approx 1000$, the density parameter of dark energy $\Omega_{de}$ is approximately $10^{-9}$ (assuming a standard $\Lambda$CDM or scalar field dark energy model). If we further assume $(1+w_{de}) \sim 10^{-1}$, Eq.~(\ref{Eqn: ce2 of equivalent DE}) implies that the effective speed of sound of the unified dark sector must satisfy $c_{e}^{2} < 10^{-10}$ to prevent the equivalent dark energy component from becoming superluminous. 
As the universe evolves toward the present epoch, where $\Omega_{m} \approx 0.3$ and $\Omega_{de} \approx 0.7$, the constraint relaxes somewhat but remains significant. Assuming $(1+w_{de}) \approx 0.1$, Eq.~(\ref{Eqn: ce2 of equivalent DE}) requires $c_{e}^{2} < 0.18$ today. It is important to note that as $w_{de}$ approaches $-1$, the required value for $c_{e}^{2}$ must become even smaller to maintain subluminality for the equivalent dark energy component. These stringent constraints suggest that for any unified dark sector model to be physically viable across all cosmological epochs, its Lagrangian $L(X,\phi)$ must inherently produce a vanishingly small effective speed of sound.

In summary, for a unified dark sector model based on a single scalar field $L(X,\phi)$ to successfully mimic a universe containing two distinct components, namely, pressureless dark matter ($c_{e(m)}^{2} = w_{m} = 0$) and subluminal dark energy ($c_{e(de)}^{2} < 1$), the effective speed of sound $c_{e}^{2}$ must undergo a significant evolution. Specifically, it must transition from a vanishingly small value (approximately $10^{-10}$ at $z \approx 1000$) during the matter-dominated epoch to a value of approximately $10^{-1}$ or lower at the present epoch. Consequently, a time-varying effective speed of sound is a fundamental requirement for most unified dark sector models. The only exception to this requirement occurs in specialized scenarios where $p_{de} = -\rho_{de}$ and $c_{e(m)} \neq 0$, such as the specific model proposed in Ref.~\cite{Sahni:2015hbf}, where the dark energy component mimics a cosmological constant.

\section{Summary and Conclusions}\label{Sec: conclusions}
In order to determine the rate of expansion of the Universe using the Friedmann equations, it is required to know how the background pressure $\bar{p}$ of the matter content is related to its energy density $\bar{\rho}$. Similarly, in order to determine the evolution of metric perturbations about the FRW background universe, a relation between the pressure perturbation $\delta p$ and the energy density perturbation $\delta \rho$ must be established. While the equation of state parameter $w$ relates the background pressure to the energy density, it is the speed of sound that relates the perturbations in pressure to those in the energy density.

For a single barotropic fluid, the speed of sound is related to the equation of state parameter, and in every gauge of cosmological perturbations, $\delta p$ is proportional to $\delta \rho$. The ratio $\delta p/\delta \rho$ for barotropic perfect fluids is the square of the adiabatic speed of sound, $c_{a}^{2} = \dot{\bar{p}}/\dot{\bar{\rho}}$. This implies that for barotropic perfect fluids, the uniform energy density gauge coincides with the gauge in which pressure is uniform. However, this is not always true in the case of single canonical or non-canonical scalar fields, where, in general, $\delta p$ is not proportional to $\delta \rho$ in every gauge. Consequently, for scalar fields, the uniform energy density gauge generally does not coincide with the gauge in which pressure is uniform. In fact, since both $\delta p$ and $\delta \rho$ are gauge-dependent quantities, their ratio depends on the choice of gauge. Moreover, this ratio at each length scale of perturbations depends on the values of $\delta p$ and $\delta \rho$, which makes it a length-dependent (or equivalently $k$-dependent) quantity.

The speed of sound of a matter component (be it a barotropic perfect fluid, a canonical scalar field, or a non-canonical scalar field) is an intrinsic property of that matter; therefore, it must depend only on the physical nature of the matter content and not on the scale of perturbation or the choice of gauge. Hence, defining the square of the speed of sound simply as the ratio of $\delta p$ to $\delta \rho$ is not meaningful in all cases.

In this paper, we consider the fact that the relationship between the variables describing the perturbations of the matter content, such as $\delta p$, $\delta \rho$, and $\delta q$, must be independent of the choice of gauge\footnote{$\delta q$ is defined in Eq.~(\ref{Eqn: Defnition delta T 0i})}. In the case of a single matter component, this requirement leads to a specific linear relation between these variables, namely Eq.~(\ref{Eqn: SF relation delta rho p and q}). One of the coefficients in this equation is identified as the square of the effective speed of sound, $c_e^2$, which is not only gauge-invariant but also independent of the spatial coordinates. It is a purely background-dependent quantity determined by the Lagrangian describing the matter content. This is referred as the ``effective speed of sound" to distinguish it from the adiabatic speed of sound. Only in the cases of barotropic perfect fluids or pure-kinetic non-canonical scalar fields does this effective speed of sound coincide with the adiabatic speed of sound. It is with this effective speed of sound that the gauge-invariant Bardeen potential $\Phi$ and the curvature perturbation $\mathcal{R}$ propagate at small scales where $k \gg aH/c_e$.

We further extend this consideration of the gauge independence of the relationship between variables such as $\delta p$, $\delta \rho$, and $\delta q$ to the case of multi-field/fluid systems, assuming all matter components are minimally coupled. In such multi-component systems, isocurvature perturbation terms also contribute to the total pressure perturbation, leading to a gauge-independent relation between the total $\delta p$, $\delta \rho$, $\delta q$, and the isocurvature perturbations. This relation is described in Eq.~(\ref{Eqn: relation delta rho p and q multi filed/fluid}) and, as in the single-component case, one of the coefficients in this equation is identified as the square of the effective speed of sound for the total system. 
The expression for $c_{e}^{2}$ given in Eq.~(\ref{Eqn: ce2 total in multi field/fluid}) is the primary result of this paper. It is shown that the gauge-invariant Bardeen potential and the curvature perturbation propagate at this speed on scales much smaller than the sound horizon. This effective speed of sound generalizes the definition introduced by Garriga and Mukhanov in Ref.~\cite{Garriga:1999vw} to a system of multiple minimally coupled non-canonical scalar fields. As a practical example, the expression for $c_{e}^{2}$ in the case of two non-canonical scalar fields is provided in Eq.~(\ref{Eqn: c_e^2 for two NC scalar fields}).

In multi-field/fluid cosmological systems, it is also possible to define a space-dependent and momentum-dependent effective speed of sound, as introduced in the literature~\cite{Romano:2018frb,Rodrguez:2020hot,Romano:2023bzn}. This quantity is related to the effective speed of sound defined in this paper, specifically Eq.~(\ref{Eqn: ce2 total in multi field/fluid}), through the expression given in Eq.~(\ref{Eqn: ce and vs reln}). It is shown that, at small scales, the effective speed of sound for a multi-filed system defined in this paper coincides with the space-dependent effective speed of sound defined in Ref.~\cite{Romano:2018frb}.

While the space-dependent and momentum-dependent effective speeds of sound may be advantageous for describing perturbations in more general cosmological systems, such as those consisting of non-minimally coupled matter contents or in modified gravity models, in this paper, we have introduced a new definition of the effective speed of sound for universes containing multiple minimally coupled components, such as canonical scalar fields, non-canonical scalar fields, and barotropic perfect fluids. 
This definition follows naturally from the requirement that the relationship between the variables defining the matter perturbations, namely,  $\delta \rho$, $\delta p$, and $\delta q$, must be independent of the choice of gauge, thereby ensuring that the effective speed of sound is a gauge-invariant quantity. Furthermore, it is shown that the effective speed of sound defined in this paper depends solely on time. It provides a gauge-independent definition for sound speed of cosmological perturbations in both single-field and multi-field/fluid systems that are minimally coupled. 
As a purely background-dependent quantity, it can be treated as a parameter to characterize cosmological perturbations in such multi-component systems. This effective speed of sound exactly coincides with the adiabatic speed of sound if the matter content of the universe consists exclusively of barotropic perfect fluids or pure-kinetic non-canonical scalar fields.

Further, in this paper, we derived a closed set of equations to describe the evolution of cosmological perturbations in a multi-field/fluid universe consisting of minimally coupled barotropic perfect fluids, and canonical and non-canonical scalar fields. The effective speed of sound of this total system serves as a key parameter in the equation governing the evolution of the gauge-invariant Bardeen potential $\Phi$. From the closed set of equations describing the evolution of cosmological perturbations, we find that:
\begin{itemize}

  \item When the multi-field/fluid system consists only of barotropic perfect fluids or pure-kinetic non-canonical scalar fields at super-horizon scales ($k \rightarrow 0$), the quantities $\delta s_{ij}/(\bar{\rho} + \bar{p})$ are conserved, where $\delta s_{ij}$ is defined as in Eq.~(\ref{Eqn: delta s ij}).
      
  \item In a general multi-field/fluid system, if the perturbations are initially adiabatic at super-horizon scales ($k \rightarrow 0$), they remain adiabatic throughout the evolution of the universe at those scales~\cite{Wands:2000dp,Gordon:2000hv,Weinberg:2003sw}. In such a scenario, the evolution of cosmological perturbations in a multi-field/fluid system is identical to that of a single barotropic perfect fluid in the $k \rightarrow 0$ limit.
\end{itemize}

Finally, we have shown that in a unified dark sector model, the effective speed of sound $c_{e}$ must evolve with time to ensure that its dark energy component possesses a subluminal perturbation speed, i.e., $c_{de}^2 \leq 1$. In fact, a unified dark sector model with a constant effective speed of sound would inevitably result in a superluminal speed of sound for its dark energy part, except in the specific case where $w_{de} = -1$, as described in the model in Ref.~\cite{Sahni:2015hbf}.

\section*{Acknowledgments}

I thank the Inter-University Centre for Astronomy and Astrophysics (IUCAA), Pune, India, for providing support through the visiting associateship program. I also acknowledge the Centre for Theoretical Physics at St. Stephen's College for the use of their research facilities.

\bibliography{ref1}

\end{document}